\documentclass[letterpaper,twocolumn,10pt]{article}
\usepackage{usenix-2020-09}
\usepackage{times}
\usepackage{tikz}
\usepackage{amsmath}
\usepackage{booktabs}
\usepackage{caption}
\usepackage{makecell}
\usepackage{sandlab}
\usepackage{pifont}
\usepackage{multirow, multicol}
\usepackage{tablefootnote}
\usepackage{subfig}
\usepackage{arydshln}

\makeatletter
\patchcmd{\maketitle}
  {\@maketitle}
  {\vspace{-0in}\@maketitle\vspace{-0in}} 
  {}
  {}
\makeatother

\newcommand{\para}[1]{{\vspace{1pt} \bf \noindent #1 \hspace{5pt}}}

\newcommand\adv{{$\mathcal{A}$}}
\newcommand\tar{{$\mathcal{U}$}}
\newcommand\pov{{PoV}}
\newcommand\typeI{{Original telemetry}}
\newcommand\typeII{{Observed telemetry}}
\newcommand\typeIII{{Rendered handpose}}

\newenvironment{packed_itemize}{
\begin{list}{\labelitemi}{\leftmargin=0.5em}
  \setlength{\itemsep}{3pt}
  \setlength{\parskip}{0pt}
  \setlength{\parsep}{0pt}
  \setlength{\headsep}{0pt}
  \setlength{\topskip}{0pt}
  \setlength{\topmargin}{0pt}
  \setlength{\topsep}{0pt}
  \setlength{\partopsep}{0pt}
}{\end{list}}

\begin{document}
\date{}

\title{\Large \bf  Can Virtual Reality Protect Users from Keystroke Inference
  Attacks?} 

\author{{\rm Zhuolin Yang}\\
University of Chicago
\and
{\rm Zain Sarwar}\\
University of Chicago
\and
{\rm Iris Hwang}\\
University of Chicago
\and
{\rm Ronik Bhaskar}\\
University of Chicago
\and
{\rm Ben Y. Zhao}\\
University of Chicago
\and
{\rm Haitao Zheng}\\
University of Chicago
} 

\maketitle
\pagenumbering{gobble}

\begin{abstract}
  Virtual Reality (VR) has gained popularity by providing immersive and interactive experiences without geographical
  limitations. It also provides a
  sense of personal privacy through physical separation.  In this paper, we
  show that despite assumptions of enhanced privacy, VR is unable to shield
  its users from side-channel attacks that steal private
  information.  Ironically, this vulnerability arises from VR's greatest strength, its immersive and
  interactive nature. We demonstrate this by designing and
  implementing a new set of keystroke inference attacks in shared virtual environments, where an
  attacker (VR user) can recover the content typed by another VR user
  by observing their avatar.  While the avatar displays noisy telemetry
  of the user's hand motion, an intelligent attacker can use that
  data to recognize typed keys and reconstruct typed content,
  without knowing the keyboard layout or gathering labeled data.
  We evaluate the proposed attacks using IRB-approved user studies across
  multiple VR scenarios. For 13 out of 15 tested users, our attacks accurately recognize 86\%-98\% of typed keys, and the recovered content retains up to 98\% of the meaning of the original typed content. We also discuss potential defenses.
\end{abstract}

\begin{figure*}[t]
  \centering
  \includegraphics[width=0.92\linewidth]{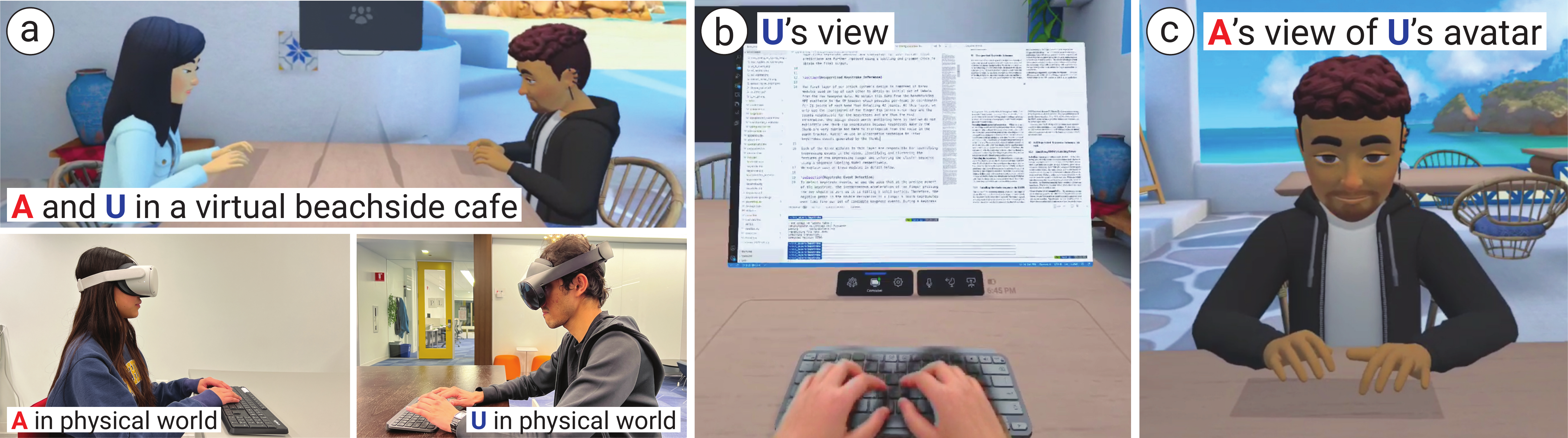}
  \vspace{-0.05in}
    \caption{An illustrative example of keystroke inference attacks in the shared VR space. (a) In this scenario, {\bf U} is replying to emails while enjoying the immersive experience of a virtual beachside cafe.  {\bf U} types comfortably and swiftly on a physical keyboard that is wirelessly connected to the VR
      headset. This is a typical setup of a VR office. Adversary
      {\bf A} is another VR user in the same virtual beachside cafe. (b) From their VR headset,  {\bf U} sees the typed content on a virtual screen and a rendered view of their own hands. (c)  {\bf A}'s view of {\bf U}'s avatar, displayed by {\bf A}'s VR headset. Using this information, {\bf A} seeks to recover the content that {\bf U} is typing. Credit: The VR scenes are screenshots taken when running the Horizon Workroom application from Meta~\cite{horizon-workrooms}.}
    \label{fig:vr_office}
    \vspace{-0.2in}
\end{figure*}

\vspace{-0.1in}
\section{Introduction}
\vspace{-0.05in}

Virtual Reality (VR) offers a whole new dimension of online interaction significantly more immersive and lifelike than those offered by existing services. By eliminating the need for physical proximity and formal attire, VR cultivates a more convenient, comfortable, and efficient environment for work, play, and social interaction. Thus, VR is increasingly adopted for both personal and professional purposes~\cite{meister2021companies,pwc2022survey}. In particular, remote work through VR has witnessed remarkable growth. Numerous companies actively promote its adoption with their remote work platforms, including Meta's Horizon Workrooms~\cite{horizon-workrooms} and VR applications and startups such as Valve~\cite{keyboardvrhtc} and vSpatial~\cite{vspatial}. Apple's recent launch of Vision Pro~\cite{visionpro} serves as further validation of the substantial interest and investment being made in VR-based remote work.

Because users interact with others using their digital representations or avatars, VR also provides them with a unique sense of privacy and added security. Representation through their digital selves can be particularly attractive for sensitive professional/work activities~\cite{giaretta2022security}. In general, VR users can customize how they present themselves and choose what they want to show of their physical environments. This ``natural filter'' can help prevent the unwanted leakage of private information without the user's knowledge. For example, it might block traditional side-channel attacks, where an attacker can use physical observations of a user or their devices to steal information, e.g., screen peeking, shoulder surfing, and keystroke inference~\cite{zhuang2005, blind1,yangusenix2023}.

In this paper, we challenge this assumption by investigating the feasibility
of  keystroke inference attacks in shared VR environments, where one VR user may attempt to recover the content typed by another user by recording and analyzing their avatar's actions. Since typing is the primary input method for sensitive information to computing systems~\cite{yangton2022},  successful keystroke inference attacks can cause significant damage, such as unauthorized disclosure of confidential or classified data that may lead to breaches of personal privacy or even financial losses. Compared to physical attacks, keystroke inference attacks over VR can be much more powerful, because a single attacker can target numerous remote victims without physical proximity constraints. 

Ironically, we show that VR's most appealing properties, its immersive and
interactive nature, are the source of VR users' vulnerability to keystroke
inference attacks.  Specifically, immersive experiences in VR rely on each
individual's {\em expressive} avatars~\cite{chu2020expressive, howard2019avatars}, which display the user's body movements, hand gestures, and facial expressions. When an individual performs physical actions such as
grasping objects, clapping hands, or typing on a keyboard, their avatar
demonstrates an approximate, digital version of those hand
movements, which are visible to other users in the same virtual space. Thus,
VR actually ``facilitates'' side-channel attacks by eliminating the need and
costs of physical sensing.

Our work shows that, while the accuracy, resolution, and granularity of
these VR movements are far from those of physical observations (e.g., a physical video
recording of the hands), intelligent attackers can still use them to extract detailed information ``carried'' by the movements, e.g., keys being typed.  Therefore, the same feature of VR that drives growing
usage and adoption is also enabling a real-world threat against user privacy
and information security.

In this paper, we describe the design, implementation, and evaluation of
keystroke inference attacks in shared VR spaces.  Figure~\ref{fig:vr_office}
plots an illustrative scenario of the attack. This attack occurs entirely in the virtual space between avatars, without any sensing or data from the physical world.  An attacker in the same virtual space as the target gathers the noisy digital representation of hand movements displayed by the target's avatar, and uses it to reconstruct the typed data. We hereby refer to these digital movements as {\bf telemetry} of the VR user.

An effective attack faces two unique challenges: (i) the excessive amount of
noise and inconsistency in the telemetry data, and (ii) the lack of knowledge
and labeled data on the target user and their physical devices, as they are
physically isolated from the adversary.  These factors, combined with the
inherent complexity and variability of human typing, make it challenging to
extract meaningful information from the available data. 
Commonly used transferability-based attacks (i.e., training a model on known
users and applying it to others) are ineffective because an individual's
typing behaviors are unique~\cite{banerjee2012biometric,
  keystrokchi15,Feit2016_how_we_type} and do not transfer to
others~\cite{yangusenix2023}.

We address these challenges by developing a componentized, self-supervised
learning pipeline. We decompose the difficult task of keystroke inference
into three sequential components, responsible for detecting keystroking events, 
identifying finger used, and recognizing the key, respectively. For each
component, we first run statistical analysis on simplified telemetry data
to produce initial inference results, leveraging general understandings of
human typing.  Then in the second component, we use these results to curate labeled training data
and train a DNN classifier (i.e., a transformer), one that learns the
intricate relationship between keystrokes and the full telemetry data. The
output of this DNN classifier is then fed to the final component.  Using
this componentized design, our attack can progressively extract, filter, and
learn important keystroke information from the noisy telemetry data, while
minimizing the spread of error in the pipeline.

Our attack also supports flexible placement of the adversary in the virtual space, whether it is sitting at the same table (Figure~\ref{fig:vr_office}), across the room, or floating in the air on a balloon.  This is achieved by applying a coordinate transformation to normalize telemetry data seen by {\em arbitrary} VR users onto a unified coordinate system so that a wide variety of attacks can be executed using a single attack pipeline.

Our work makes four key contributions. 
\begin{packed_itemize} 
\vspace{-0.05in}

\item We identify three forms of keystroke inference attacks that can be
  executed by a VR adversary in the same virtual space as the target user.  These attacks consider different levels of information access available to the adversary, including the original 3D telemetry collected by the target's headset, the transformed 3D telemetry sent to the adversary to render the target's avatar, and its 2D version extracted directly from the adversary's screenshots of the avatar (i.e., Figure~\ref{fig:vr_office}(c)).

\item We design an effective attack pipeline to infer keystroke content from
  noisy attack data, using a componentized
  self-supervised learning approach.

\item We perform IRB-approved user studies to evaluate the proposed attacks under a
  variety of VR scenarios and settings, varying avatar distance and
  placement, keyboard device, and typing content. Moreover, we test the attacks on 15 users. On average, our attacks accurately recognize 91.3\% of the typed
  keys, and the recovered content retains 71.5\% of the meaning of the typed content (see Table \ref{tab:all_participants}).

\item We explore potential defenses and find that adding zero-mean Gaussian
  noise to the telemetry data can reduce attack effectiveness while
  preserving the immersive experience to some degree.  

\vspace{-0.05in}
\end{packed_itemize}

To the best of our knowledge, our work is the first to explore and
demonstrate the feasibility of keystroke inference attacks in shared VR
environments. Our work identifies the crucial challenge of designing VR
systems that can provide immersive and engaging experiences while ensuring
personal and information security. Also, our attack methodology represents a substantial departure from the conventional understanding of transformers. We hope our results can inspire more self-supervised, transformer-based attack/defense designs.
\vspace{-0.0in}
\section{Background and Related Work}
\vspace{-0.05in}
In this section, we first describe existing keystroke inference attacks in both physical and VR environments. We then discuss the key features adopted by VR to create immersive experiences (which are the base of our proposed attack) and the general landscape of privacy threats in VR. 

 \vspace{-0.1in}
 \subsection{Existing Keystroke Inference Attacks}
 \vspace{-0.05in}
 \label{subsec:existingattacks}
In a keystroke inference attack, an attacker can reconstruct a target's typed content without control over their input devices. We divide existing  attacks into three categories:  (i) physical attacks that place sensors near the target to capture their typing actions,  (ii) remote attacks analyzing the target's network traffic,  and (iii) attacks where the target is a VR user. 

\para{Physical attacks.} By physically placing sensors (e.g., cameras,
microphones, inertial measurement units (IMU), electromagnetic (EM),
  and radio frequency (RF) devices) near a target, attackers can record data
related to the target's typing actions. The recorded sensor data includes
audio~\cite{zhuang2005} and video~\cite{clearshot,lim2020revisiting,beware,blind1,ispy,seeingdouble}, as
well as vibration, EM~\cite{emccs21}, WiFi and LTE
measurements~\cite{spidermon, windtalker,wifi2018}. As pressing different
keys causes changes in signals recorded by sensors, attackers can use
statistical or machine learning analysis to determine keystroke content. For example, a well-known audio-based attack~\cite{zhuang2005} placed a
microphone next to the target to capture key-specific sounds generated by
typing on a mechanical keyboard, and trained a machine learning model to
identify key entries from the recording. Follow-up works replace microphones
with other sensors like IMUs, WiFi/LTE radios, or EM sniffers. Other
examples use cameras to capture the screen and the pressing fingertip
reflected by the target's eyes or eyeglasses~\cite{ispy, seeingdouble}, or to record the reflections around the pressing fingertip produced by a reflective typing surface, such as a tablet~\cite{blind1}. 

Physical attacks are mostly ineffective in the VR attack settings where the adversary, a VR user, cannot place or access sensors or side channels (e.g., eye reflection or surface reflection) in the target's physical space.  Moreover, many physical attacks require knowledge of the keyboard and its layout, which are inaccessible to the VR attacker. 

\para{Remote attacks by traffic analysis.}  Song et al.~\cite{song2001timing} show that by analyzing encrypted SSH traffic, attackers can detect when a user is typing a password, and estimate the password length and content. This is possible because SSH pads data for encryption using an 8-byte boundary and sends packets upon each keystroke in interactive mode. Thus, attackers can infer password length by counting packets and estimate its content using packet inter-arrival times. On the other hand, this attack is only applicable to SSH sessions. 

\para{Attacks against VR users.}  Recent works have examined keystroke inference attacks against VR users~\cite{lingcns2019, Arafat2021, Luo2022, Meteriz2022}. 
They all target VR users who use a virtual keyboard to input text.  To enter a key, the user may either rotate their head to move the cursor on the virtual keyboard and make a specific hand gesture to select the key~\cite{lingcns2019}, or type by moving two fingers (one per hand) in the air~\cite{Meteriz2022} or by pointing handheld controllers~\cite{Arafat2021}. These virtual keystrokes result in simple and slow typing movements that can be monitored by attackers using methods such as deploying a stereo camera~\cite{lingcns2019} or a pair of WiFi devices~\cite{Arafat2021} near the target,  or by installing malware directly on the target's VR headset~\cite{lingcns2019, Meteriz2022}.  These attacks also assume perfect knowledge of the ``virtual keyboard'' layout and screen location, and often labeled data from the target. 

\vspace{-0.1in}
\subsection{Creating Immersive Experiences in VR} 
\label{subsec:immersive}
\vspace{-0.05in}

\para{Expressive avatars.} The most compelling aspect of VR is its capacity to generate immersive experiences through expressive avatars. Expressive avatars foster a heightened sense of presence and connection by closely emulating users' real-life movements and expressions. To build such avatars, modern VR systems (e.g., Meta Quest 2/3/Pro, HTC VIVE XR Elite) deploy real-time hand tracking and isomorphic rendering, so users can see and manipulate virtual objects with their own hands. The rendering of hand movements and gestures in their avatars adds an additional layer of immersion, allowing users to communicate, gesture, and interact with others naturally in the VR environment. This level of realism and expressiveness provides a true sense of presence and social interaction~\cite{QoMEX2020}.

\para{Hand tracking.} Hand tracking in VR tracks the position and orientation of a user's physical hands. Today, it is achieved using multiple cameras and IMUs deployed on the user's VR headset~\cite{occulusimu}. The accuracy of hand tracking is influenced by various factors such as the relative position of the hands and headset, the number and placement of IMUs and cameras, lighting conditions in the environment, and the type and speed of hand movements. Even with advanced VR systems, some level of error, latency, or jitter in tracking may still occur, particularly when the user's hands are moving rapidly.

Currently, the accuracy of hand tracking falls short of the level required for precise keystroke recognition~\cite{han2020megatrack}. This is due to the fast and intricate finger movements that are specific to each user and constantly changing, along with the natural obstruction of the headset-mounted cameras by the user's hands. To overcome this challenge, commercial VR systems utilize a physical keyboard as the input device, while researchers suggest incorporating additional sensors, such as IMUs on the user's wrists~\cite{tapid2021}, to improve the capture of finger-tapping movements.

\vspace{-0.05in}
\subsection{Privacy Threats and Defenses in VR}
\vspace{-0.05in}
As VR systems can gather large amounts of data, including user activity, location, and behavior,  they are inherently susceptible to privacy attacks that steal or misuse personal information or sensitive data.  A recent study~\cite{garrido2023sok} summarizes the landscape of VR privacy threats and defenses by reviewing 68 publications, which lists 30 VR attacks, including the two keystroke inference attacks~\cite{lingcns2019, Arafat2021} discussed in \S\ref{subsec:existingattacks}. The other 28 attacks focus on revealing the user's demographics (e.g., age, gender, ethnicity) or sensitive attributes (e.g., emotion, physical and mental health, and wealth). 

\vspace{-0.1in}
\section{Keystroke Inference Attacks in a Shared Virtual Environment}
\label{sec:threat}
\vspace{-0.02in}

We study a new, VR-based scenario for keystroke inference attacks where both the target and the
adversary are VR users in the same virtual environment (see Figure~\ref{fig:vr_office}). While the target types,  the adversary seeks to recover the typed content using information displayed by the target's avatar. Without the constraints of physical proximity to the target, a single adversary can attack many remote victims, increasing the potential reach and impact of the attack.  Next, we discuss the motivation behind the attack and the threat model. 

\vspace{-0.1in}
\subsection{Motivation and Real-World Implications}
Our attack is driven by two important trends in VR. {\em First}, productive typing on a full physical keyboard is becoming a common mode of input in VR systems. Compared to virtual keyboards, physical keyboards offer precision, comfort, familiar use, and haptic feedback~\cite{vrphykeyboard}. Popular VR platforms like Oculus Quest 2/3/Pro and HTC VIVE and VR apps like vSpatial and Immersed already support physical keyboards through Bluetooth connection~\cite{keyboardvr, keyboardvrhtc, vspatial, immersed}. Many VR users also prefer to use them to input text, especially when working for extended period of time~\cite{vrphykeyboard}.
As users can now type efficiently and professionally in their VR workspace, text input in VR will quickly grow to encompass  emails, documents, and other text-based materials that may contain sensitive information and personal data.

{\em Second}, VR enables immersive interaction among people in shared virtual environments where each VR user is represented by their own avatar rather than their physical self.   This contrasts with the well-known privacy issue in physical spaces where people are directly visible to each other.

These two developments prompt the following question, which motivated our study:  {\em \textbf{``Is it safe to type sensitive information in a shared virtual environment, since VR users can only see each other's avatars?''}}

\begin{figure}[t!]
    \centering
      \includegraphics[width=1\linewidth]{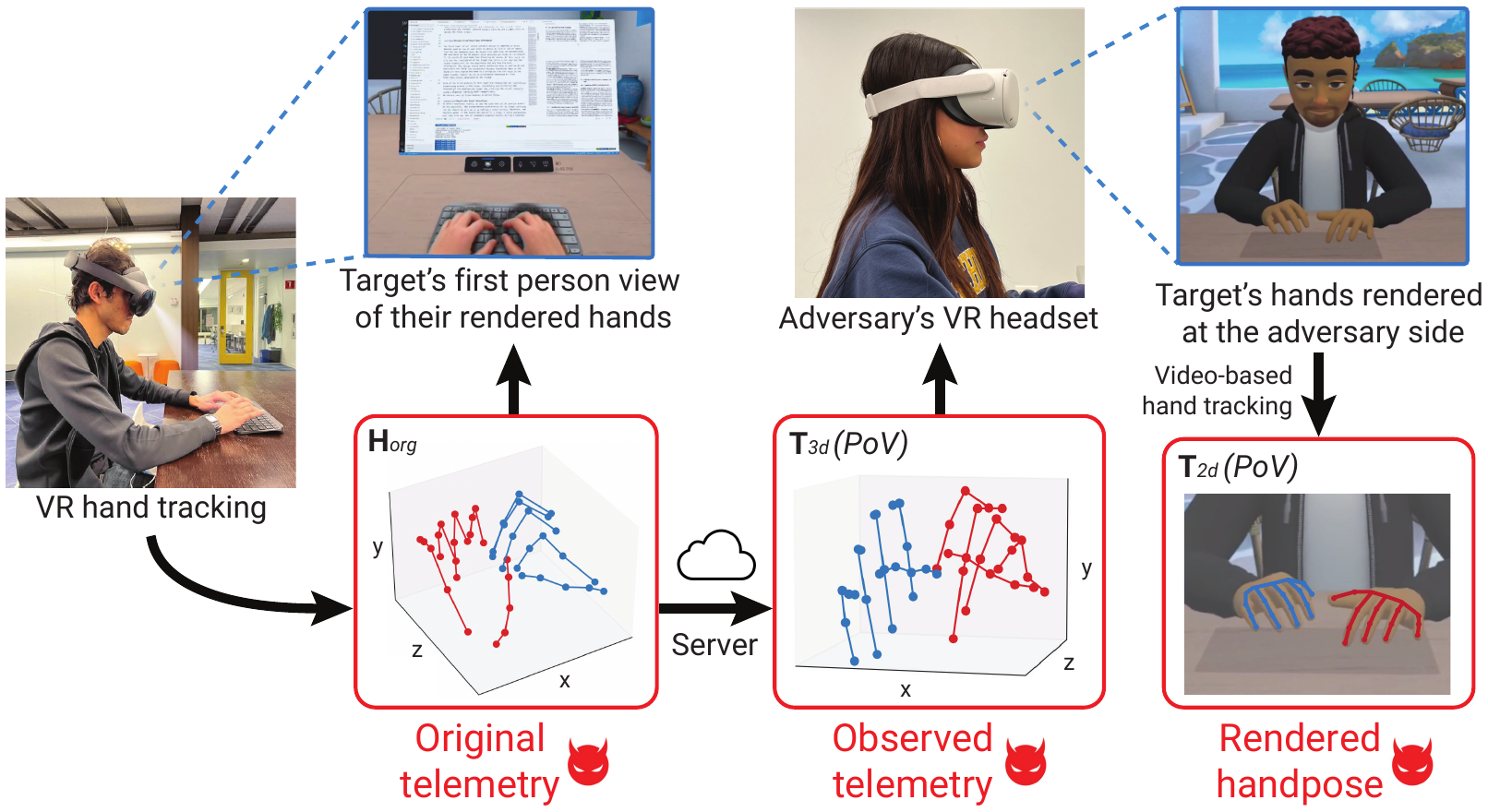}  
      \vspace{-0.15in}
      \caption{The three keystroke inference attacks considered by our work, operating on the original telemetry, the observed telemetry (used to render the target's avatar),  or the video of the target's avatar displayed on the adversary's VR screen.} 
      \label{fig:handpose_formats}
      \vspace{-0.15in}
\end{figure}

\vspace{-0.05in}
\subsection{Threat Model}   
\vspace{-0.05in}
\label{sec:threeattacks}

We consider a VR user \tar{},  who is doing work in VR by typing on a physical keyboard wirelessly connected to their VR headset. We consider an adversary \adv{} who is  another VR user sharing the virtual environment with \tar. To enable seamless interaction, the VR platform uses an expressive {\em avatar} to represent each user in the virtual environment, which displays their nonverbal behaviors such as body movements and hand gestures \cite{howard2019avatars, chu2020expressive}. 

We assume that \adv{} can {\em only gather data related to \tar's avatar}. To render  \tar's avatar to be seen by \adv, the headset of \adv{} will receive a continuous stream of \tar's telemetry data, i.e., \tar's handposes. We note that although hand tracking is a built-in feature of modern VR headsets, the regulations governing the access and sharing of its data are still in development. Upon user consent, VR apps can access, store, and stream this information to render avatars on other devices. Therefore, we investigate various attack scenarios by varying the adversary's level of information access and the design of the VR app. Following this process, we identify three versions of the attack,  defined by the telemetry data gathered by the adversary (see Figure~\ref{fig:handpose_formats}). 

\begin{packed_itemize}
\vspace{-0.08in}

\item {\bf Attack I: \typeI{} attack operating on $\mathbf{H}_{org}$}. This represents the strongest adversary, one who has access to either the target's headset or the VR rendering server~\cite{garrido2023sok}. Here $\mathbf{H}_{org}$ is the 3D telemetry data on \tar's hands collected by \tar's headset, which is extracted from a top-down (or bird's-eye) view of the hands. 

\item {\bf Attack II: \typeII{} attack operating on $\mathbf{T}_{3d}(\pov)$}. Here the adversary obtains \tar's avatar data by setting up a virtual camera. To render the virtual hands of \tar's avatar to be seen by \adv,  the VR system needs to transform the original telemetry data  $\mathbf{H}_{org}$ to screen coordinates of \adv's virtual camera. The latter is defined by $\pov$, \adv's camera point-of-view (PoV) with respect to \tar. For example, when \adv's avatar (thus the camera)  directly faces \tar, i.e., $\pov=(0,0)$, the observed telemetry $\mathbf{T}_{3d}(0,0)$ captures a frontal-view of \tar's hands. Depending on the VR system/app design, $\mathbf{T}_{3d}(\pov)$ can be calculated by the VR server and sent to \adv's headset, or by \adv's headset after receiving $\mathbf{H}_{org}$.  Here we note that the latter makes it possible for \adv{} to gather $\mathbf{H}_{org}$ directly from \adv's VR device and launch the Attack I mentioned above.

\item { \bf Attack III: \typeIII{} attack using $\mathbf{T}_{2d}(\pov)$}. The attack operates on a 2D projection of $\mathbf{T}_{3d}(\pov)$ used by Attack II such that the depth-to-camera data is removed. This represents a limited adversary who has no access to any telemetry data, but instead records \tar's avatar shown on \adv's VR screen and applies a hand tracking tool to extract the 2D handpose per video frame.  \vspace{-0.08in} 
\end{packed_itemize}

In our attack design, the only assumption is that the attacker is aware of the language used by the target (English in this study). More importantly, the attacker has {\bf no} other information about the target. This means the attacker lacks knowledge of the target's keyboard layout and position, does not have access to labeled data or prior observations of the target, and does not have any additional side-channel information besides the avatar telemetry/handpose data previously discussed.

\para{Extension to virtual keyboards.} Note that our proposed attack naturally extends to future VR systems allowing controller-free virtual typing on a surface. The attack works because it requires only handpose data and no knowledge of a ``keyboard.'' While such virtual typing is infeasible today due to insufficient accuracy in hand tracking, it may become a reality in the near future as VR hardware/technology advances. Ironically, improvements in the accuracy of VR hand tracking systems will only increase the potency of our attack. 
\vspace{-0.05in}
\section{Design Challenges and Baseline Solutions}
\label{sec:noise}
\vspace{-0.05in}
The proposed attack scenario is unique because the adversary only utilizes telemetry/handpose data of the target's avatar. However, this also presents significant difficulties in constructing a successful attack due to the high levels of noise present in the data. In this section, we address this challenge through a thorough analysis of the attack data and an examination of potential attack designs. Our discussion focuses on the original telemetry attack ($\mathbf{H}_{org}$) because $\mathbf{T}_{3d}$ and $\mathbf{T}_{2d}$ are lossy transformations of $\mathbf{H}_{org}$. 

\vspace{-0.05in}
\subsection{A Closer Look at the Telemetry Data}
\label{subsec:noisy_handpose_data}
\vspace{-0.05in}

\begin{figure*}[t!]
  \centering
    \includegraphics[width=0.92\textwidth]{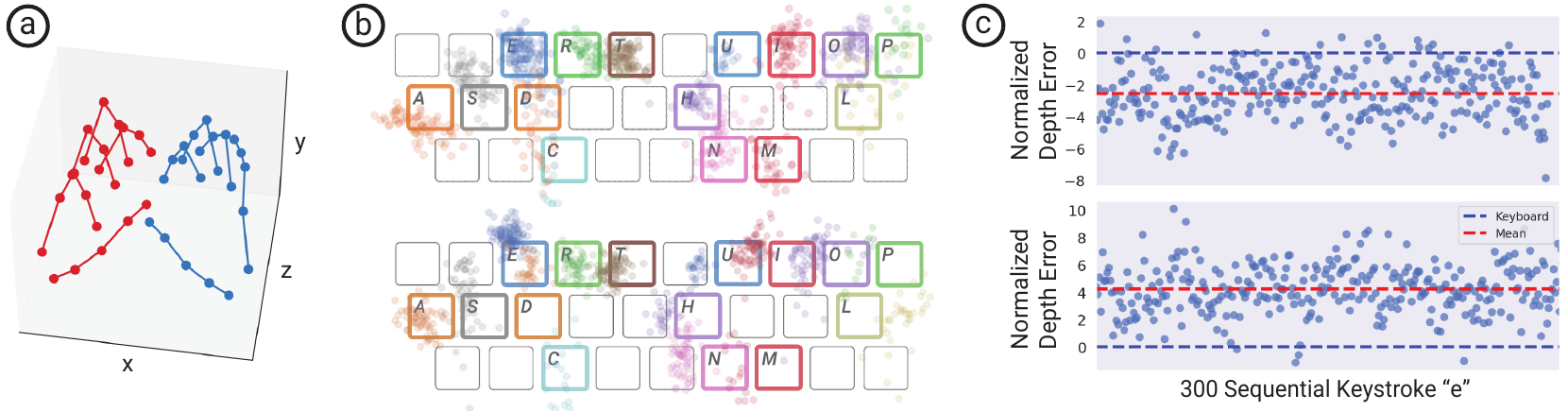} 
    \vspace{-0.125in}
    \caption{(a) A sample instance of $\mathbf{H}_{org}$ where the left pinky is pressing a key. (b) VR collected $(x, z)$ touchpoints from 2 participants. The corresponding ground truth key positions are marked with the same color. Top-15 frequent letter samples are shown. (c) Depth of sequential keystrokes when `e' is typed by Participants 1 and 2. Normalized depth error puts the ground truth at zero and scales by the height of an individual key. Standard Deviation for Participant 1: 1.639. Standard Deviation for Participant 2: 1.853.} 
    \label{fig:xyznoise}
  \vspace{-0.2in}
\end{figure*}

We examine the 3D telemetry data $\mathbf{H}_{org}$ produced by the Oculus Quest Pro headset, using bird's-eye views of the wearer's hands captured by the headset's four cameras. At a sampling time $t$, the 3D telemetry is represented by $(x, y, z)$ for each of the 46 joints on both hands (23 joints per hand).  This 3D coordinate system is arranged by $p$, the physical position of the headset measured at the start of the VR session, and remains unchanged throughout the session even as the user's head moves around. Therefore, given the $(x, y, z)$ coordinate of a joint, $y$ is the depth of the joint to $p$, while the $(x, z)$ are the deviation from $p$ in the horizontal plane. Figure~\ref{fig:xyznoise} (a) shows a sample instance of $\mathbf{H}_{org}$, where the left pinky is pressing a key. Since the physical keyboard is placed on a horizontal plane, the $y$ value captures the relative {\em height} to the keyboard while the $xz$ values indicate the {\em positions within} the keyboard.

\begin{figure}[t]
   \centering
     \includegraphics[width=0.7\linewidth]{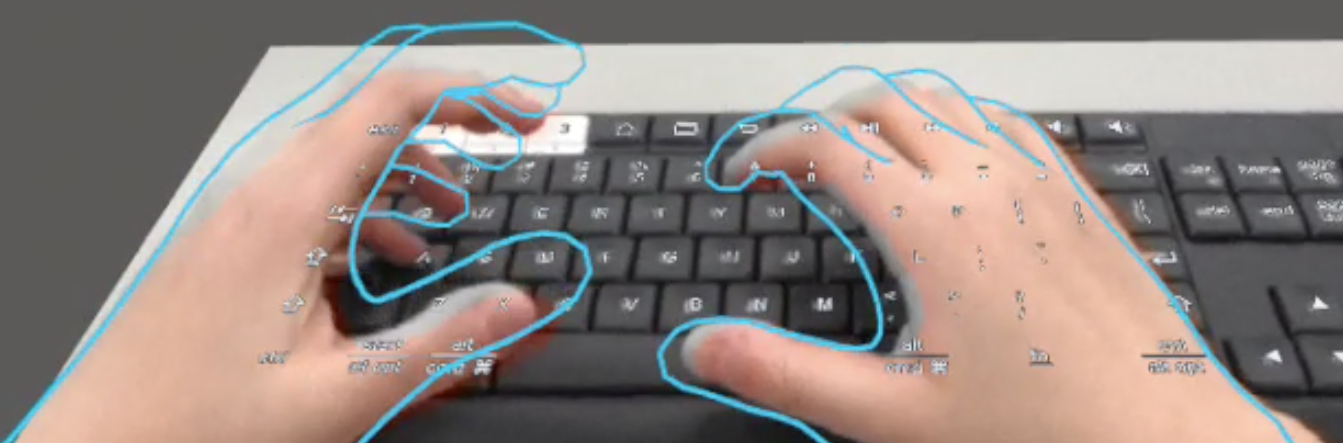} 
     \vspace{-0.05in}
     \caption{The virtual hands rendered by VR headset, overlaid on top of their corresponding physical hands.}
     \label{fig:handposeonhand}
     \vspace{-0.2in}
 \end{figure}

\para{Virtual vs. physical hands.} Figure~\ref{fig:handposeonhand} plots a sample observation seen by \tar{}, where the {\em virtual} hands (rendered by \tar's headset from $\mathbf{H}_{org}$) are overlaid on top of the physical hands. The virtual hands do not align with the physical hands, especially at the fingertips, demonstrating that the telemetry data is noisy. This is expected as hands can often obstruct fingertips from the view of the headset cameras and typing on a physical keyboard is rapid and delicate. 

\para{Error analysis.} We present an in-depth analysis of the noise in the telemetry data. As it is challenging to obtain ground truth positions of hand joints, we choose to analyze the fingertip coordinates when the user is pressing a key. Accurate recording of timing and keystrokes is ensured by the physical keyboard. For each keystroking event, we determine the corresponding pressing fingertip by manually reviewing an external video of the typing hands.

{\bf {\em Noise in $(x,z)$}:} \hspace{4pt} Figure \ref{fig:xyznoise} (b) shows, for two users, the spread of $(x,z)$ values of pressing fingertips (reported by $\mathbf{H}_{org}$) on top of the actual physical keyboard. Each press point is marked by the color of the key being pressed. For clarity, we only show the top 15 frequently used letters. Both users exhibit clear deviations, with variations among keys and users. The physical size and spacing of keys do result in some level of separation in the $xz$-plane of pressing points.

{\bf {\em Noise in $y$}:}  \hspace{4pt} The noise effect on the depth $y$ value is greater due to the tendency of users to rest their hands on the physical keyboard and only slightly lift their fingers to press keys. This results in a small depth difference between pressing and non-pressing fingertips. 
Figure~\ref{fig:xyznoise} (c) plots, for two users, the error of depth measurements for pressing the letter `e', normalized by the key height (4.4mm). These results come from 300 instances of  typing `e', extracted from a random typing session. For each instance, we find the pressing fingertip's $y$ value, subtract from it the ground truth depth of the keyboard, and divide the residual by the key height. Ideally, the values should be very close to $0$ (i.e., the blue line marking the ground truth depth of the keyboard). The actual results demonstrate significant deviations and display a user-specific bias. For both users, over 50\% of the typing instances have depth errors higher than $2\times$key height.

The same depth noise affects non-pressing fingers, making it challenging to identify both keystroking events and their corresponding pressing fingertips. To study the impact on identifying the pressing fingertip, we compute, for each keystroking event, the deviation from the ``lowest'' non-pressing fingertip's $y$ to the pressing fingertip's $y$. Ideally, the deviation should be positive and $\geq 1\times$ key height, since the pressing fingertip is the lowest across all fingers. Yet we find that, for both users, {\bf more than 30\%} of the instances have negative values, i.e., at least one non-pressing fingertip is wrongly reported as pressing lower than the pressing fingertip. 

\para{Summary.} These findings collectively demonstrate the presence and impact of excessive noise in the 3D telemetry data $\mathbf{H}_{org}$, particularly for the $y$ values.  We also note that, for the other two forms of handpose data ($\mathbf{T}_{3d}$ and $\mathbf{T}_{2d}$), the errors are exacerbated by the lossy transformations from $\mathbf{H}_{org}$. 

\vspace{-0.08in}
\subsection{Exploring Attack Design Options}
\label{subsec:explore}
In our quest to find an effective attack,  we attempted the following methods, but none proved to be successful.  We present them below and summarize key limitations of each. 

\para{Denoising telemetry data.} An important question is whether techniques such as data pre-processing and denoising can enhance the quality of the telemetry data. Unfortunately, this is challenging as the noise in the telemetry data follows a Gaussian distribution. For instance,  noise in the depth $y$ value of the pressing fingertip is confirmed to be Gaussian through KS goodness-of-fit tests, with a $p$-value greater than 0.5.

\para{Manual inspection.} A curious attacker can manually inspect each telemetry frame to identify keystroke events and the corresponding pressed key. However, we attempted this and found that the random noise in the telemetry data makes it challenging for the human eye to accurately identify keystrokes and the finger responsible. 

\para{Attack by transferability.}  Another option is to leverage transferability across users~\cite{Meteriz2022}, i.e., collecting accurately labeled keystroke/telemetry data from a set of users, training a DNN inference model, and applying it to infer keystroke content of other target users.  We examined this attack among 7 users, applying leave-one-out cross validation.  In each round, we selected 6 users, trained a supervised DNN keystroke classification model using labeled telemetry data from them, and tested the trained model on the one new user.  We evaluated the attack by comparing the typed and recovered content (after apply spell correction).  The results were consistent:  the attack on a new user unseen during training produced a large character error rate (56\% $\pm$11\%) and could not recover any meaningful content.  But on trained users, the models can correctly detect and recognize more than 90\% of keystrokes (i.e., $<$10\% character error rate). This aligns with other keystroke studies, which show that human typing behaviors are user-specific and transferability-based attacks are ineffective~\cite{Meteriz2022,yangusenix2023}. 

\para{Statistical analysis of fingertip motion.}  With no labeled data from the target to train an ML model, one can apply unsupervised inference by studying the target's fingertip movements. Specifically, one can extract fingertip positions from the telemetry data, and apply statistical analysis to detect the keystroking events and identify the corresponding pressing fingertip. However, given the abundant noise in the telemetry data, the attack is ineffective (see our experiments in \S\ref{subsec:comparison}).

Our finding aligns with existing studies on VR keystroke recognition~\cite{tapid2021}: the depth $y$ of the VR hand tracking lacks the required accuracy to identify keystroking events and the pressing finger. To compensate for these errors, researchers propose placing extra IMUs on the user's wrists and collecting labeled data to train a DNN model for finger identification~\cite{tapid2021}. Yet both are infeasible under our attack scenario.

\begin{figure*}[t!]
  \centering
    \includegraphics[width=0.95\linewidth]{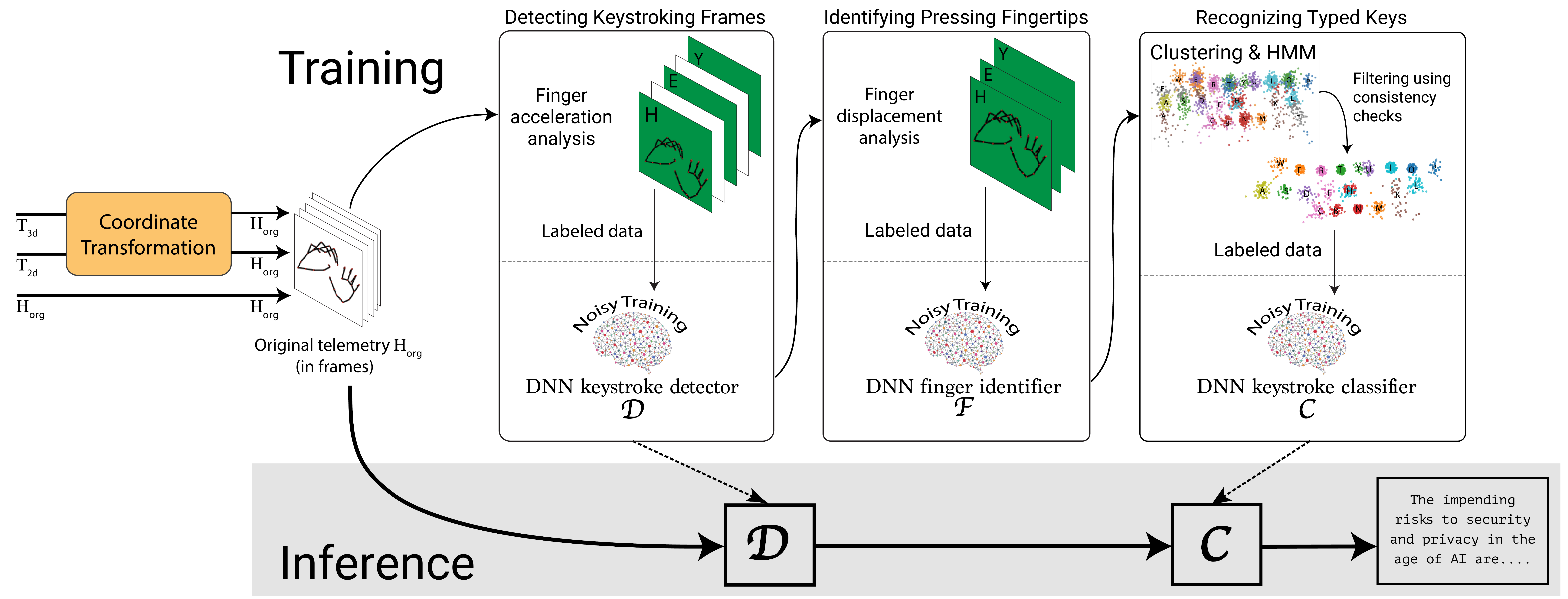}  
    \vspace{-0.05in}
    \caption{The end-to-end attack design, using a componentized self-supervised learning pipeline} 
    \label{fig:attackpipeline}
  \vspace{-0.2in}
\end{figure*}

\vspace{-0.1in}
\section{Our Proposed Attack}
\vspace{-0.06in}
\label{sec:attack}
In this section, we present the proposed attack design. We begin with the key insights behind our design and a summary of the attack pipeline. Later in \S\ref{sec:details} we delve into the individual components and the implementation of the attack. 

\vspace{-0.08in}
\subsection{Design Insights}
\vspace{-0.06in}
\label{sec:designInsights}
Our exploration of the transferability-based attacks (in \S\ref{subsec:explore}) revealed that, with accurately labeled training data from the target user, a sophisticated DNN model can be trained to learn the intricate connections between keystrokes and noisy 3D telemetry data, allowing it to correctly recognize over 90\% of keystrokes. This suggests that an effective attack is possible if we can self-label some of the target's 3D telemetry data and use them for model training.  Such labeling is done on the target's data since transferability-based attacks are ineffective.

The concept of self-curating labeled training data from the target data is an established technique in self-supervised learning~\cite{amini2022self, zhu2005semi}. A recent work~\cite{yangusenix2023} applies this concept to recognize keystrokes from a video recording of the target's physical hands. This is done by first applying hand tracking to locate the target's fingertips in each video frame, followed by a HMM-based statistical analysis to estimate labels for a small subset of the video frames.  These labeled data are then used to train CNN models to recognize the keystrokes captured by all video frames. 

But when applying the method of~\cite{yangusenix2023} to our attack scenario, we observe significant failures (see Table \ref{tab:all_participants}).  This is because the amount of noise included in our attack data is higher than that of~\cite{yangusenix2023}, i.e., our attack data is the noisy hand tracking results estimated by the VR headset, while \cite{yangusenix2023} uses the raw video capturing the target's physical fingers as the input to their CNNs. 
This qualitative change renders the method of ~\cite{yangusenix2023} ineffective and requires a new approach. 
  
\para{Self-supervised learning of noisy handposes using task-specific transformers.}  To handle the extremely noisy telemetry data, we design our attack based on two key insights. First, different from CNN models, the self-attention mechanism of transformers enables the model to learn dependencies between hand joints and give more importance to the joints responsible for performing a specific task, e.g., pressing a key on the keyboard. As such, transformers can be highly effective in extracting key patterns from noisy handpose data and greatly boost attack robustness in noisy data settings. 

Second, despite the common belief that transformers require vast amounts of training data,  our work demonstrates their suitability for our attack, even with limited data. This is attributed to two design decisions. The first is that we exclusively train transformers on the current attack data, ``forcing'' them to concentrate solely on learning crucial features from this specific dataset. Next, we break down the main task into three sub-tasks or components, and train transformers separately for each component. This empowers each transformer to focus on its designated learning task. By adopting these design choices, we effectively leverage transformers to execute our attack, even when faced with limited training data.

\vspace{-0.09in}
\subsection{Attack Overview}
\label{subsec:attackoverview}

Based on the above insights, we propose a new attack design, which  applies a set of three sequential components to infer typed characters from noisy telemetry data. Our design tackles the challenging issue of noisy telemetry data through a two-step self-supervised learning approach, which is applied to each of the three components separately. By applying this self-learning method to each individual component, we design the attack to progressively extract and learn important keystroke information from the noisy input, while effectively reducing the spread of error in the end-to-end attack pipeline. 

\para{The three components.} We break down the main task into three
  components: detecting keystroking telemetry frames, identifying pressing fingertips, and recognizing typed keys. The first component focuses on detecting telemetry frames\footnote{Here each telemetry frame refers to the telemetry data collected on the target user at a specific time $t$.} that
  represent keystroking events. The next component takes as input the
  detected keystroking frames and identifies the specific finger that
  touches the keyboard. The third component leverages the detected finger
  information to locate the pressing location on the keyboard and analyzes
  the pressing locations of all the keystroking frames to estimate the
  sequence of keys that the target user has entered.

\para{Training per-component transformers.}  For each component, we first
self-curate labels for its input telemetry data by performing statistical and
motion analysis on the data.  Here the statistical analysis focuses on a
subset of hand joints closer to the keyboard and leverages general
understandings of human typing behavior.  After creating the initial labels
for the telemetry data, we use them to train a deep neural network (DNN)
model dedicated for this component. Note that this DNN model learns the
detailed relationship between keystrokes and telemetry data from 3D
coordinates of {\em all} hand joints (i.e., the full telemetry).  To
minimize the impact of the noise present in both the self-generated labels
and the telemetry data, we employ multiple noise-aware training strategies
during DNN model training.

Here we choose transformers as the DNN architecture for the first two
components, but CNN for the last component due to the issue of class
imbalance (details in \S\ref{sec:details}). 

\para{End-to-end attack pipeline.} Figure~\ref{fig:attackpipeline} plots the attack pipeline, which takes as input a sequence of 3D telemetry frames ($\mathbf{H}_{org}$) captured during the target's typing session, and applies the three sequential components to analyze the data and train DNN models.   
This pipeline produces three DNN models, one per component: (i) a transformer based keystroke detector $\mathcal{D}$, which identifies the set of telemetry frames where the target enters a key, (ii) a transformer based finger identification $\mathcal{F}$, which determines the finger used to press the key, and (iii) a CNN-based keystroke classifier $\mathcal{C}$,  which predicts the key entered for each keystroking telemetry frame.  The final inference process only employs $\mathcal{D}$ and $\mathcal{C}$, since $\mathcal{F}$ is only used to produce cleaner labels for the last component and train a more effective $\mathcal{C}$. 

\para{Supporting \MakeLowercase{\typeII{}} and \MakeLowercase{\typeIII{}} attacks.} To enable these attacks (i.e., using $\mathbf{T}_{3d}$ or $\mathbf{T}_{2d}$), we deploy a pre-processing module that transforms $\mathbf{T}_{3d}$ or $\mathbf{T}_{2d}$ into a coordinate system equivalent to that of $\mathbf{H}_{org}$. This is the ``coordinate transformation'' module in Figure~\ref{fig:attackpipeline}. 

\vspace{-0.08in} 
\section{Detailed Attack Design}
\vspace{-0.08in} 
\label{sec:details} 

We now present in detail each of the three sequential components of the attack pipeline, the pre-processing module, followed by the end-to-end attack implementation.  

\vspace{-0.08in} 
\subsection{Detecting Keystroking Frames}
\label{sec:detectingKeys}
The first component studies the input sequence of 3D telemetry frames to identify keystroking events and their associated telemetry frames.  This is done in two steps, first applying statistical and motion analysis to self-label the frames,then using these labeled data to train a transformer-based DNN model, leveraging noise-aware model training techniques. Note that our discussion assumes the telemetry frame is from $\mathbf{H}_{org}$. 

\para{Step 1: Self-curating labels via motion analysis.}  When typing on a keyboard, each keystroking action is a characteristic movement of the finger, which starts from reaching the key,  then a brief moment of zero acceleration as the finger hits the key, followed by a return to the starting position. This movement creates a distinctive pattern, represented as a positive peak in the plot of the second derivative of the keystroking finger's depth coordinate (or the $y$ value of the fingertip in the telemetry data). 

This general pattern only stands when the fingertip depth ($y$) values are measured at a very high accuracy (i.e., sub-mm precision). Today's VR hand tracking is far from that. The noise in the telemetry data, in combination with hesitant typing behavior, produces many ``fake'' positive peaks. On the other hand, these fake peaks often display a smaller amplitude than the real ones.  Thus, we model the overall peak series as a mixture of two Gaussian distributions~\cite{yangusenix2023} and compute a decision threshold to produce an equal error rate between admitting a fake keystroke and missing an actual keystroke. 

 \para{Step 2: Training a transformer-based keystroke detector.} In Step 1, we analyze the unique feature of fingertip acceleration in the $y$ axis to identify keystroke events, resulting in a set of ``rough'' labels for all telemetry frames. Using these labeled data, we train a {\em transformer}-based~\cite{vaswani2017attention} keystroke detection model in Step 2.  The choice of the transformer architecture deserves some explanation. The telemetry data consists of 3D positions of 42 hand joints in each frame, which have natural correlations in their movements and configurations across time and space. Furthermore, these joint configurations have spatial constraints due to the hinge nature of hand joints. To effectively learn and accurately predict finger activity, it is essential to have a model that understands these correlations and constraints.  Along this line, the self-attention mechanism used by transformers~\cite{bahdanau2014neural} is well-suited for our learning task as it allows the model to learn dependencies between hand joints and give more importance to the joints relevant for performing a particular task (i.e., keystroke). 

 To model the spatial-temporal relationship among hand joints, we choose the DSTA-Net architecture proposed by Shi et al.~\cite{dstanet_accv2020} for skeleton-based gesture recognition. This architecture calculates spatial attention to create a combined spatial representation of joints and then utilizes it to compute the temporal attention of joints across time.  The pre-trained model provided by~\cite{dstanet_accv2020} and its training data only cover a single hand. Thus, we adapt the model architecture to support both hands and train the detection model from scratch to run a binary classification task (keystroke or non-keystroke). The input to the transformer-based classifier is a collection of 16 telemetry frames to capture the spatial-temporal movement of the hands. Specifically,  we first identify each telemetry frame labeled as ``keystroke'' in Step 1,  then take 7 telemetry frames before it and 8 telemetry frames after it to form a telemetry segment of 16 frames,  and  label it  as ``keystroke.''  We also build non-keystroke segments by grouping 16 non-keystroke telemetry frames between two detected keystroke events.  

 \para{Noise-aware model training.}  Since both the telemetry data and the labels we self-curate (during Step 1) contain errors, we apply two noise-aware training techniques when training the transformer model.   First, we apply mixup data augmentation~\cite{zhang2017mixup} to prevent the model from overfitting into the training data and label.  This is done by generating new training samples via blending two existing training samples using convex combinations, teaching the model to make linear predictions on those new data. The presence of noise in both the data and labels impedes the model's ability to learn these linear relationships and naturally prevents overfitting.  

 Second, we use bootstrapping for refurbishing labels which relies on the idea that as training progresses, the model learns to predict the correct class with higher confidence. Therefore, for a sample with a noisy label, the model's own prediction is a better source of the ground truth label and should be used as such during training whereas correctly labeled samples should be trained in the regular way. To identify the noisy labels, Arazo et al.~\cite{arazo2019unsupervised} leverage the idea that noisy labels are learned in the later stages of training. Therefore, the difference in the loss can be used to identify the noisy labels which can then be trained using the model's predictions. In practice, convex combinations of the noisy and model-predicted labels are used. For samples that are noisy with high probability, the model-predicted labels are weighted higher in the refurbished label and vice versa.

 \vspace{-0.08in}
 \subsection{Detecting the Pressing Fingertip}
 \label{sec:identifyFinger}
 \vspace{-0.06in}
 \para{Self-curating labels using finger displacement.} After detecting keystroke events, it is important to identify the finger used to generate each keystroke in order to estimate the location of the keystroke. For this, we adopt the idea from~\cite{yangusenix2023} that given any position of the hand, the finger used to press the key needs to move the most in comparison to the non-pressing fingers. Therefore, we compute the displacement of each fingertip, at the moment of a keystroke, from its mean position across the entire typing session. We label the fingertip with the maximum displacement as the pressing finger. 
 However, due to subtle typing behaviors and sensor noise, finger identification with this method is prone to errors thus some pressing fingertips are often incorrectly labeled. 

 \para{Training a transformer for fingertip identification.} Using the labeled fingertips, we train another transformer model to leverage features in the raw telemetry data. 
 Again, this model is trained in a noise-aware manner from scratch on a set of 16 raw frames for each detected keystroke and predicts which finger was used to press a key. After training, this model is evaluated on the same data used for training in order to correctly predict the noisy labels. However, noisy training has its limits. If the proportion of noisy labels in the training set is beyond a certain threshold, noisy training techniques are unable to produce a reliable model due to overfitting. We identify these cases by observing the clustering output (\S\ref{sec:hmm}) and avoid using the transformer in this component if it produces a worse clustering result. 

\vspace{-0.08in}
\subsection{Recognizing Typed Keys}
\label{sec:hmm} \vspace{-0.06in}
  
  \para{Self-curating labels using clustering \& HMM. } After obtaining fingertip labels (from the fingertip transformer), we use the 3D coordinates of each identified fingertip, at the precise moment of the keystroke, to estimate a touchpoint map which corresponds to the layout of the keyboard. Since keys are separated and have fixed locations on a keyboard, this form factor naturally clusters keystrokes generated from the same key and allows us to run unsupervised inference to reconstruct the typed content. We use K-Means clustering with K=38 to cluster the keystrokes and label these clusters arbitrarily. This labeling gives us a sequence of cluster IDs where each element in the sequence corresponds to a keystroke in the order that it occurred. The final task is to map each of these elements to one of 29 keys (period, comma, space key, and the 26 letters of the English alphabet). This is the kind of task Hidden Markov Models are well-suited to solve. We explain the detailed HMM implementation in Appendix~\ref{appendix:hmm}. The output of the HMM gives us another set of labels which we use to train a DNN model for keystroke classification. 
  
  Due to the accumulated errors in our pipeline, there are errors in the HMM's prediction of the typed content. Therefore, we use the consistency checks between the clustering and the HMM's output introduced by~\cite{yangusenix2023} to identify the incorrectly recognized typed keys. As a result, instead of solely relying on noisy training, we can remove noisy labels from the self-curated labels to create a cleaner training dataset for the DNN-based keystroke classifier. The details of these consistency checks along with the explanation of other clustering related components are explained in Appendix~\ref{appendix:clusteringdetails}.

  \para{Training a CNN for recognizing keys.} The natural choice for training the keystroke classifier would be the transformer architecture used previously. However, after filtering,  the training dataset (of 29 classes) is much smaller compared to those used to train the previous transformer models. This dataset is also severely class imbalanced due to the uneven distribution of the alphabet frequency in the English language. Due to these factors, a transformer-based keystroke classifier does not perform well.
  
  Thus, we opt for a 3D-CNN model (ResNext-101~\cite{xie2017aggregated}) for keystroke classification. This model is pretrained on the Jester~\cite{materzynska2019jester} dataset which contains videos of humans performing different gestures and can therefore be fine-tuned with limited data.  However, our data is in the form of 3D hand coordinates whereas the 3D-CNN is trained on images. Following common practice~\cite{sharma2019deepinsight, zhu2021converting}, we convert our hand joint data to images by using the $(x, z)$ coordinates as the pixel locations and the $(x, y, z)$ coordinate as the pixel value. Like before, such a sample in the training dataset consists of 16 consecutive raw telemetry frames corresponding to a detected keystroke and its label obtained from the HMM.

\vspace{-0.1in}
\subsection{Coordinate Transformation}
The adversary with access to \MakeLowercase{\typeI{}} data ($\mathbf{H}_{org}$) poses the strongest threat, as they have the ``cleanest'' data among all three attacks.  For the other two attacks, $\mathbf{T}_{3d}(\pov)$ is a lossy, transformed version of $\mathbf{H}_{org}$,  while $\mathbf{T}_{2d}(\pov)$ is a projected version of $\mathbf{T}_{3d}(\pov)$ with the depth dimension removed.  Since the above described pipeline is designed using the 3D coordinate system of $\mathbf{H}_{org}$, we execute the other two attacks by first applying a coordinate transformation, which converts their attack data into an equivalent version of $\mathbf{H}_{org}$. 

\para{Converting $\mathbf{T}_{3d}(\pov)$ to $\mathbf{H}_{org}$.} $\mathbf{T}_{3d}(\pov)$ is the telemetry observed from \adv{}'s point of view. It is generated from $\mathbf{H}_{org}$ by first applying 3D transformation defined by $\pov$, including translation, scaling, and rotation~\cite{rotationmatrix}. This is followed by perspective projection~\cite{rotationmatrix, perspectiveprojection} where the telemetry joints are projected onto a 2D viewing window (like a camera frame) based on \adv{}'s $\pov$. Together, this transformation and projection create  misalignments and distortions that interfere with the motion analysis used by our attack pipeline. 

To execute the observed telemetry attack, we apply a 3D rotation to $\mathbf{T}_{3d}(\pov)$ by calculating the dot product of each telemetry joint's coordinates\footnote{Since $z$ is in a different unit than $x$ and $y$, normalizing them into the same scale is needed before applying the 3D rotation.} with standard rotation matrices~\cite{rotationmatrix}. This allows us to convert $\mathbf{T}_{3d}(\pov)$ into an $\mathbf{H}_{org}$ approximation, which can be input to our pipeline. 

\para{Multi-camera \MakeLowercase{\typeIII{}} attack.}
$\mathbf{T}_{2d}(\pov)$ is a 2D version of $\mathbf{T}_{3d}(\pov)$ without the $z$ (i.e., the depth information). Given the hefty information loss, the attack becomes ineffective if  the adversary \adv{} has only a single $\mathbf{T}_{2d}(\pov_{1})$. However, \adv{} can easily compensate for the information loss by adding another {\em hidden} camera at a different $\pov_2$ and collect $\mathbf{T}_{2d}(\pov_{2})$, and use them to recover the lost depth values based on  stereo epipolar geometry~\cite{stereoDepthEstimation}. Setting up multiple, simultaneous, {\em hidden} cameras by a single VR user is a functionality supported by VR systems and developments. For instance, \adv{} could register multiple sybil accounts and place their avatars in the same VR environment with \tar{} to capture  $\mathbf{T}_{2d}$ from different angles.

\vspace{-0.1in}
  \subsection{Putting It All Together}
  \label{subsec:attackpipeline}
  Once the DNN models are trained, we run inference using the DNN-based keystroke detector $\mathcal{D}$ and classifier $\mathcal{C}$. The keystroke detector identifies the telemetry frames which contain keystrokes and these are passed to the classifier to infer the key used for that keystroke. The finger identifying DNN becomes redundant in the inference stage as its purpose is to obtain a cleaner clustering that improves the labels in the training set of the keystroke classifier. The keystroke classifier sidesteps the issue of finger identification by directly inferring the keystroke labels using patterns in raw telemetry frames. 

  After reconstructing the typed content, we further improve it using spell check. We find that conventional spell checking tools operate at a word level whereas spell checking can benefit from context. Thus, we use the spelling and grammar checking tool available in Google Docs, which makes use of the context using machine learning techniques~\cite{googledocs}, to improve the recovered text. We build an automated version of the Google checker using the Selenium~\cite{SeleniumHQ} library.

\vspace{-0.05in}
\section{Evaluation}
\vspace{-0.05in}
\label{sec:eval}
We assess the effectiveness of the proposed attack through user studies approved by our Institutional Review Board (IRB21-1396). We evaluate five key aspects of the attack: 

\begin{packed_itemize}\vspace{-0.08in}

\item {\bf Effectiveness of \MakeLowercase{\typeI{}}, \MakeLowercase{\typeII{}}, and \MakeLowercase{\typeIII{}} attacks (\S\ref{subsec:evaldata})}, which utilize three types of telemetry data; 

\item {\bf Performance under different VR configurations (\S\ref{subsec:evalscenario})}, where we vary the distance between the avatars, the keyboard device, and the content type; 

\item {\bf Attack effectiveness on 15 users} of varying typing styles (\S\ref{subsec:evaluser});

\item {\bf Comparison to other solutions (\S\ref{subsec:comparison})}, including the baseline solutions (attack by transferability and unsupervised inference) discussed in \S\ref{subsec:explore} and the attack design of \cite{yangusenix2023}; 

\item {\bf Attack complexity (\S\ref{subsec:evaltime})} in terms of computation time on a standard server and a breakdown across components. 
\vspace{-0.1in}
\end{packed_itemize}

\vspace{-0.15in}
\subsection{Experimental Setup}
\vspace{-0.05in}
\para{User (or target) configuration.} By default, each of our study participants wears a Meta Quest Pro VR headset and sits in front of a table (on a wheeled chair).  A physical keyboard (Logitech K375s) is placed on top of the table and connects to the headset via Bluetooth.  Before each typing experiment, we leave sufficient time so that our participants can freely adjust their headset, the chair, and the keyboard. On average, each typing session lasts about 25 minutes.

While wearing the VR headset, each participant types on the keyboard. Through the headset, the participant can see a rendered (virtual) display showing the text they have entered and those to be typed. They can also see their virtual hands, rendered locally using the tracked telemetry $\mathbf{H}_{org}$. By default, we ask each participant to type, on average, 26 email sentences randomly selected from a popular corporate email dataset~\cite{Enron}, and to correct their typing mistakes using the backspace key. To simulate actual working conditions, each participant is allowed (and encouraged) to take breaks during the typing session, including pausing, leaving, and returning to their seat. Both the table and the keyboard remain stationary during a typing session. 

 \begin{figure}[t]
   \centering
     \includegraphics[width=0.7\linewidth]{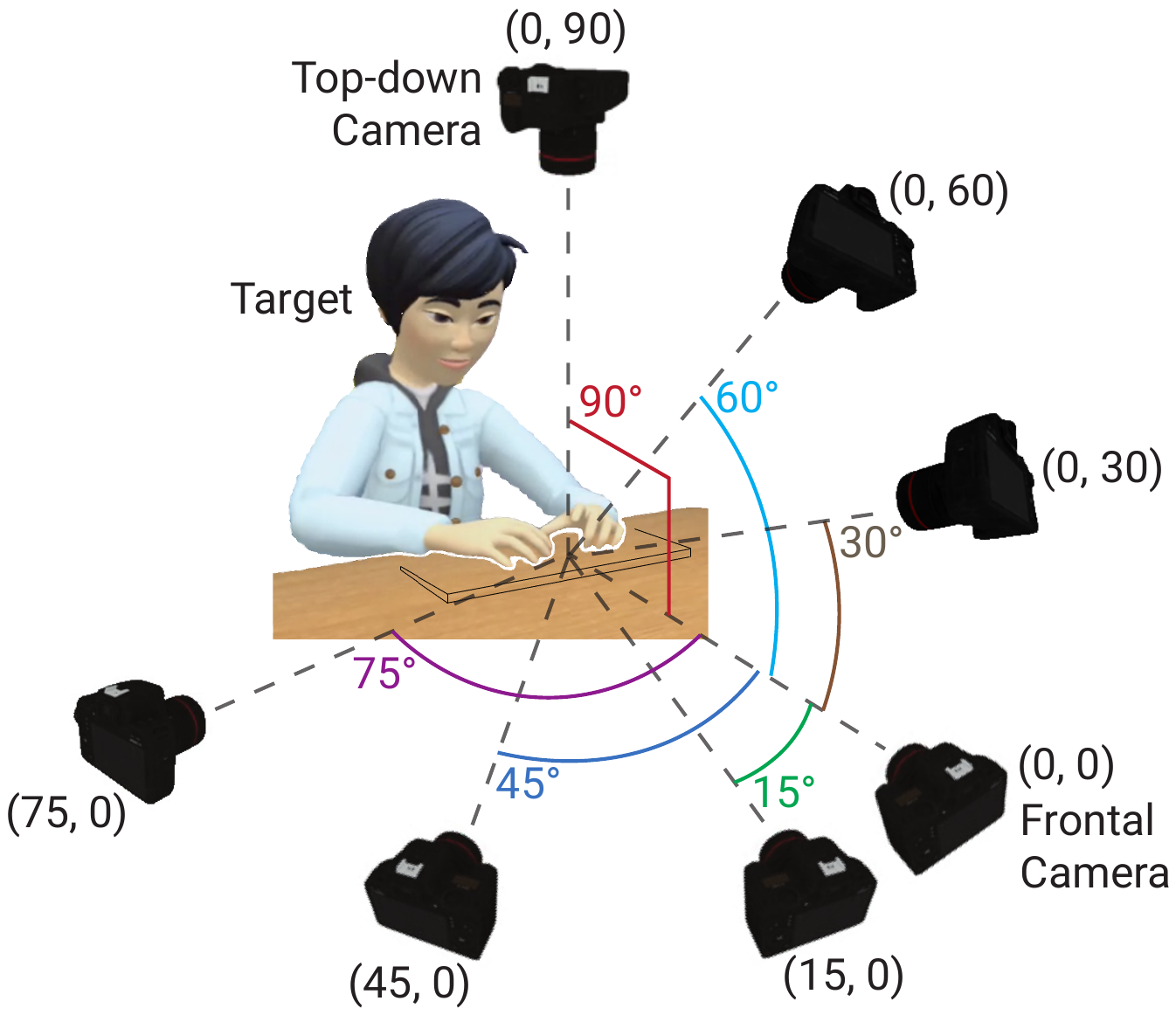} 
     \vspace{-0.05in}
     \caption{A graphical illustration of virtual camera $\pov$s. $(0, 0)$ means a frontal camera at the same height as the keyboard pointing towards the target's typing hands. $(x, y)$ means the camera is horizontally, vertically deviates by $x, y$ degrees from $(0, 0)$, e.g., $(0, 90)$ represents a top-down camera. Credit: The target avatar is created using the built-in avatar tool on a Meta Quest Pro headset.} 
     \label{fig:pov_illustrate}
     \vspace{-0.15in}
\end{figure}

\para{Attacker configuration.} We consider an attacker who is able to acquire the telemetry data of the target. The telemetry data is in one of the three forms: \MakeLowercase{\typeI{}} $\mathbf{H}_{org}$, \MakeLowercase{\typeII{}} $\mathbf{T}_{3d}(\pov)$, and \MakeLowercase{\typeIII{}} $\mathbf{T}_{2d}(\pov)$.  The sampling rate of telemetry is around $60$Hz, a standard setting used by both Quest 2 and Pro VR headsets. The latter two attacks depend on the attacker parameter $\pov$, i.e., the virtual camera placement configured by the attacker. Figure~\ref{fig:pov_illustrate} illustrates the $\pov$ configuration.  $\pov=(0, 0)$ means the attacker places a frontal virtual camera at the same height as the keyboard, which points toward the target's typing hands. $\pov=(x, y)$ means the virtual camera deviates vertically and horizontally from $(0, 0)$ by $x, y$ degrees, respectively.  For all $\pov$ configurations, the virtual camera always points toward the target's typing hands. 

\para{Dataset.} We build a Unity program to collect telemetry data used by our attack and ground truth typing input for our evaluation. For each user, we collect their telemetry data over a typing session ($\approx$500 words).  Each session's data, i.e., a set of telemetry frames, is used to implement the attack for the current user and the current session. Specifically, we first self-curate binary labels (i.e., keystroking or non-keystroking) for a subset of the telemetry frames, and use them to train a keystroke detector $\mathcal{D}$. We feed the entire session's data to $\mathcal{D}$ to identify the set of keystroking frames. Next, for all keystroking frames, we curate finger-specific labels to train a finger identifier $\mathcal{F}$ and apply $\mathcal{F}$ on all keystroking frames to help curate the key-specific labels, followed by a consistency check filter to construct the labels for a subset of keystroking frames. The resulting labeled data is used to train the keystroke classifier $\mathcal{C}$, which is then applied to all keystroking frames to produce the attack output. 

Here we note that the quantity of training data used to train the three DNNs ($\mathcal{D}$, $\mathcal{F}$, $\mathcal{C}$) differs across users and sessions. As reference, Table~\ref{tab:dataset} in the Appendix displays, for each of the 15 users, the quantity of telemetry frames used to train these DNNs and the quantity of inference data being fed into the trained DNNs.

\para{Evaluation metrics.}  Following an existing work~\cite{yangusenix2023}, we evaluate each attack attempt by comparing the typed and inferred content at the character, word, and semantic levels.

\begin{packed_itemize}
  \vspace{-0.05in}
  \item Character error rate (CER): It measures errors at the character level by computing the ratio of the number of errors and the length of the actual typed content. The number of errors is measured by the minimum edit distance, comprising insertions, deletions, and substitutions, required to convert the inferred content to the typed content. The lower the CER, the higher the accuracy of the inferred content.
  \item Word error rate (WER): The WER is computed using the same ratio as the CER with the difference that the edit distance is computed at the word level. It is worth mentioning that while WER penalizes words that are not exact matches, they may be comprehendible by a human reader. We use the WER function available within the Hugging Face~\cite{wercompute} library as it uses dynamic string alignment to best match word indices for comparison.
  \item Content similarity using plagiarism detection: Plagiarism detectors are used to compute similarity between documents and are thus a natural fit for evaluating the recovered content. We use the plagiarism detector offered by CopyLeaks~\cite{copyleaks} which provides a similarity score in the range of 0-100\% for the actual and inferred content. With moderate errors in the inferred content, the similarity score operates as a hybrid between character and word level similarity. However, with higher levels of errors, the similarity score usually collapses and is much lower than the word recovery rate (1 - WER). We refer to the obtained similarity score as the \textbf{similarity}. \vspace{-0.08in}
\end{packed_itemize}

\vspace{-0.08in}
\subsection{Attack Effectiveness vs. Telemetry Data}
\label{subsec:evaldata}

We begin by examining the attack performance on \MakeLowercase{\typeI{}}, \MakeLowercase{\typeII{}}, and \MakeLowercase{\typeIII{}} data. It is intuitive to assume that the \MakeLowercase{\typeI{}} attack is the strongest as the adversary has access to $\mathbf{H}_{org}$, the ``cleanest'' data among the three. As explained earlier  in \S\ref{subsec:attackpipeline},  the \MakeLowercase{\typeII{}} data ($\mathbf{T}_{3d}(\pov)$) is produced from $\mathbf{H}_{org}$ using a lossy, non-linear perspective projection, while the \MakeLowercase{\typeIII{}} data ($\mathbf{T}_{2d}(\pov)$) suffers even more information loss from the 3D-to-2D projection.

In Table~\ref{tab:transformed_single}, we compare the performance of the three attacks, by varying the attacker's virtual camera $\pov$. For consistency, we invited a single participant as the attack target for all the experiments.  We also include the result for the {\bf two-camera  \MakeLowercase{\typeIII{}} attack}, where the attacker sets up two virtual cameras to collect two $\mathbf{T}_{2d}$ at different $\pov$s. 

\begin{table}[t]
\renewcommand{\arraystretch}{1.005}
\centering
\setlength{\tabcolsep}{0.666em}
\begin{tabular}{|l|l|l|rrr|}

\hline

\textbf{Attack}& \textbf{Attack} & \textbf{Attacker's} &\textbf{CER} & \textbf{WER} & \textbf{Sim.} \\
\textbf{Type} &\textbf{Data} & {\bf PoV}  & (\%) & (\%) & (\%) \\
\hline

\multirow[c]{2}{*}{\parbox{1.6cm}{Original Telemetry}} & \multirow[c]{2}{*}{$\mathbf{H}_{org}$} & \multirow[c]{2}{*}{--} & \multirow[c]{2}{*}{4.7} & \multirow[c]{2}{*}{14.2} & \multirow[c]{2}{*}{87.4} \\
& & & & & \\ 
\hline 

\multirow[c]{7}{*}{\parbox{1.6cm}{Observed Telemetry}} 
& \multirow[c]{7}{*}{$\mathbf{T}_{3d}(PoV)$} 
&$(0, 0)$  & 6.6 & 21.6 & 80.4 \\
& &$(0, 30)$ & 6.2 & 21.4 & 83.7 \\
& &$(0, 60)$ & 6.5 & 21.4 & 81.5 \\
& &$(0, 90)$ & 7.2 & 23.8 & 74.6 \\
\cdashline{3-6}
& &$(15, 0)$ & 5.8 & 18.0 & 86.1 \\
& &$(45, 0)$ & 6.2 & 19.4 & 84.2 \\
& &$(75, 0)$ & 7.6 & 21.0 & 77.6 \\
\hline 

\multirow[c]{3}{*}{\parbox{1.6cm}{Single camera Rendered Handpose}} 
& \multirow[c]{3}{*}{$\mathbf{T}_{2d}(PoV)$} 
&$(0, 0)$ & 29.3 & 78.0 & 0.0 \\
& &$(0, 45)$ & 23.7 & 58.6 & 2.9 \\
\cdashline{3-6}
& &$(15, 0)$ & 28.5 & 66.8 & 0.0 \\
\hline 

\multirow[c]{5}{*}{\parbox{1.6cm}{Two-camera Rendered Handpose}} 
& \multirow[c]{5}{*}{\parbox{1cm}{\centering $\mathbf{T}_{2d}(0, 0)$ \& $\mathbf{T}_{2d}(PoV)$}}
&$(0, 15)$ & 6.9 & 20.8 & 85.3 \\
& &$(0, 45)$ & 6.3 & 20.6 & 81.7 \\
& &$(0, 75)$ & 6.5 & 20.0 & 80.5 \\
\cdashline{3-6}
& &$(15, 0)$ & 7.6 & 21.2 & 79.0 \\
& &$(30, 0)$ & 5.0 & 17.4 & 84.1 \\
\hline

\end{tabular}
\vspace{-0.05in}
\caption{Performance of \MakeLowercase{\typeI{}}, \MakeLowercase{\typeII{}} and \MakeLowercase{\typeIII{}} attacks, using $\mathbf{H}_{org}$, $\mathbf{T}_{3d}$, and $\mathbf{T}_{2d}$ handposes, respectively. We also include an extended \MakeLowercase{\typeIII{}} attack using two $\mathbf{T}_{2d}$ at different camera $\pov$s.}
\label{tab:transformed_single}
\vspace{-0.3in}
\end{table}

\para{Effectiveness of \MakeLowercase{\typeI{}} attack.}  Our results show
that the \typeI{} attack (using $\mathbf{H}_{org}$) is highly successful -- the
attacker effectively recovers 85.8\% of the words in the typed text. The
recovered text has a semantic similarity of 87.4\% to the original.

\para{\typeII{} attack with different $\pov$s.} In the \MakeLowercase{\typeII{}}
attack using $\mathbf{T}_{3d}(\pov)$, the attacker needs to set up a
virtual camera at an observation point $\pov$ to collect
$\mathbf{T}_{3d}(\pov)$.  We experiment with 7 different $\pov$ values and
find that the attack is consistently effective  -- the attackers can recover
around 78\% of the typed words.  The performance is slightly less than that
of the \MakeLowercase{\typeI{}} attack because the inverse transformation of
$\mathbf{T}_{3d}(\pov)$ produces a noisy version of $\mathbf{H}_{org}$.  We
observe the same performance trend from other participants (see
Table~\ref{tab:transformed_others} in Appendix). 

\para{Two-camera \MakeLowercase{\typeIII{}} attack.}  Using a single
virtual camera, the \MakeLowercase{\typeIII{}} attack (using
$\mathbf{T}_{2d}(\pov)$) is ineffective because $\mathbf{T}_{2d}(\pov)$ is a
noisy 2D projection of $\mathbf{H}_{org}$, making it hard to estimate the 3D
telemetry.  Yet we can overcome this challenge by setting up another virtual
camera at a sufficiently different $\pov$. With two different
$\mathbf{T}_{2d}$ observations, we can estimate the 3D telemetry
$\mathbf{H}_{org}$ based on the stereo epipolar
geometry~\cite{stereoDepthEstimation}.  In our experiments, we complement a
frontal-view camera $\mathbf{T}_{2d}(0,0)$ with either a side-view or a
vertical-view camera.  Results in Table~\ref{tab:transformed_single} show
that the two-camera \MakeLowercase{\typeIII{}} attack performs comparably to the
\MakeLowercase{\typeII{}} attack.  Again, these observations are confirmed across
other users (see Table~\ref{tab:transformed_others} in Appendix).

\para{Real-world implications.}  The above results demonstrate the
  effectiveness of the three attacks. The success of the two-camera
  \MakeLowercase{\typeIII{}} attack is particularly alarming due to two
  reasons. First, the attack is simple and stealthy. To collect
  $\mathbf{T}_{2d}(\pov)$, the attacker \adv{} does not need to hack any
  server or VR headset, but just acts as a benign user in the
  same VR environment with \tar{} and performs screen recordings
  of their VR screen on the headset. After applying video-based hand
  tracking\footnote{Open source hand tracking tools are commonly available
    online, e.g., MediaPipe~\cite{mediapipe} and OpenPose~\cite{openpose}.}
  on the recorded video of the VR view, \tar{} obtains
  $\mathbf{T}_{2d}(\pov)$.  Second, \adv{} can register multiple sybil
  accounts in VR and record \tar{}'s typing from many $\pov$s.  With more
  versions of $\mathbf{T}_{2d}$,  \adv{} can potentially obtain better estimates of the
  original telemetry  $\mathbf{H}_{org}$ and improve attack
  effectiveness. 

\vspace{-0.1in}
\subsection{Effectiveness under Different VR Settings}
\label{subsec:evalscenario}

To test the generalizability of our attack, we conduct multiple experiments by varying factors involved in our end-to-end attack pipeline. These include changing the (virtual) distance between the adversary's virtual camera to the target's avatar, the physical keyboard type, the typed content, and the amount of data available to launch the attack. To maintain consistency, a single participant in our user study was the target in all these attack scenarios.

\para{Varying avatar distance.}  To assess the impact of avatar distance in the virtual world, we conducted an experiment in which four adversaries were positioned at increasing distances (0.6m, 1.2m, 1.8m, and 2.4m) from the target avatar during a single typing session. The adversaries recorded four separate sequences of transformed 3D telemetry ($\mathbf{T}_{3d}(0, 0)$) using their virtual cameras. The results are summarized in Table \ref{tab:distance_cer}, where we consider 0.6m attack as the baseline with CER $=$ 3.8\% and report the CER difference between the 1.2m/1.8m/2.4m attack and the baseline. The results confirm that the virtual distance between the target and the adversary avatars does not impact the attack performance. 

\begin{table}[h]
\vspace{-0.05in}
\renewcommand{\arraystretch}{1.005}
\centering
\setlength{\tabcolsep}{1em}
\begin{tabular}{|l|r|r|r|}
\hline
\textbf{Avatar distance (m)} & \textbf{1.2} & \textbf{1.8} & \textbf{2.4} \\
\hline
CER $-$ CER$_{0.6}$ (\%) & $-$0.17 & $-$0.06 & $+$0.03\\
\hline
\end{tabular}\vspace{-0.05in}
\caption{Attack performance when the virtual distance between the
  target and the adversary avatars varies. The default attack is
  conducted with the avatar distance $=$ 0.6m. We report the CER
  difference from that of the default attack (CER $=$ 3.8\%).}
\label{tab:distance_cer}
\vspace{-0.15in}
\end{table}

\para{Varying keyboard type and size.}
People can use different keyboards of various sizes. To evaluate our attack under such conditions, we ask our participant to type on three different keyboards: Logitech K375s, Logitech MX Keys Mini, and Apple Magic keyboard. The results for these experiments, displayed in Table~\ref{tab:typing-device}, show that our attack is successful against all three physical keyboards. 

\begin{table}[h]
\vspace{-0.05in}
\renewcommand{\arraystretch}{1.005}
\centering
\setlength{\tabcolsep}{0.8em}
\begin{tabular}{|l|l|rrr|}
\hline
\multirow[c]{2}{*}{\textbf{Keyboard}} 
& \textbf{Dimension} & \textbf{CER} & \textbf{WER} & \textbf{Similarity} \\
& (cm) & (\%) & (\%) & (\%)\\
\hline
Logitech K375 & 42.7 $\times$ 13.0 & 4.7 & 14.2 & 87.4 \\
Logitech MX & 29.3 $\times$ 12.8 & 6.2 & 19.6 & 86.4 \\
Apple Magic & 28.0 $\times$ 11.5 & 6.6 & 17.9 & 82.4 \\
\hline
\end{tabular}
\vspace{-0.05in}
\caption{Attack performance when the target types on three different physical keyboards.}
\vspace{-0.125in}
\label{tab:typing-device}
\end{table}

\para{Varying content type and length.}
We investigated the performance of our attack with different types and lengths of text content. In addition to corporate emails, a participant typed approximately 500 words of text from scientific papers and medical patents, which contain technical terms and uncommon words. To evaluate the effect of limited typing data, we also conducted the attack when the participant typed only around 250 words. The results, presented in Table~\ref{tab:content}, indicate that our attack is effective with various content types and lengths tested. 

\begin{table}[h]
\renewcommand{\arraystretch}{1.005}
\centering
\setlength{\tabcolsep}{0.7em}
\begin{tabular}{|l|l|rrr|}
\hline
 & \textbf{Word} & \textbf{CER} & \textbf{WER} & \textbf{Similarity} \\
 & \textbf{Count} & (\%) & (\%) & (\%)\\
\hline
Abstract & 509 & 5.2 & 18.1 & 83.3 \\
\hdashline
Patent & 501 & 4.6 & 20.8 & 80.6 \\
\hdashline
\multirow[c]{2}{*}{Emails} & 501 & 4.7 & 14.2 & 87.4 \\
 & 262 & 5.4 & 17.6 & 84.5 \\
 \hdashline
 w. numbers & 488 & 5.5 & 18.7 & 81.9 \\
\hline
\end{tabular}
\vspace{-0.05in}
\caption{Attack performance when the target types content of different
  kinds and lengths.}
\vspace{-0.2in}
\label{tab:content}
\end{table}

\para{Impact of numbers in text.}
The Enron~\cite{Enron} dataset includes sentences that contain numbers, which account for 1.2\% of keystrokes. Using these sentences, we evaluate our attack to study the impact of number keys.  Intuitively, the presence of numbers in the typed content would affect the HMM-based labeling process because HMM (trained on text) is unable to recognize numbers and will label them as letter keys.  However, because these number keys are physically separated from letter keys on the keyboard, they will be filtered out by our consistency checks and will not appear in the training data for the keystroke classifier.  As such, the moderate presence of numbers has minimum impact on our attack. Our results in Table~\ref{tab:content} (last row)  also confirm this hypothesis.

\begin{table*}[t]
  \renewcommand{\arraystretch}{1.005}
  \footnotesize
  \centering
  \setlength{\tabcolsep}{.266em}
  \begin{tabular}{|l|ccc|ccc|ccc|ccc|ccc|ccc|ccc|ccc|}
  \hline 
  \multicolumn{1}{|c|}{\textbf{}} & \multicolumn{3}{c|}{\textbf{P1}} & \multicolumn{3}{c|}{\textbf{P2}} & \multicolumn{3}{c|}{\textbf{P3}} & \multicolumn{3}{c|}{\textbf{P4}} & \multicolumn{3}{c|}{\textbf{P5}} & \multicolumn{3}{c|}{\textbf{P6}} & \multicolumn{3}{c|}{\textbf{P7}} \\
  \textbf{Approach} & \textbf{CER} & \textbf{WER} & \textbf{Sim.} & \textbf{CER} & \textbf{WER} & \textbf{Sim.} & \textbf{CER} & \textbf{WER} & \textbf{Sim.} & \textbf{CER} & \textbf{WER} & \textbf{Sim.} & \textbf{CER} & \textbf{WER} & \textbf{Sim.} & \textbf{CER} & \textbf{WER} & \textbf{Sim.} & \textbf{CER} & \textbf{WER} & \textbf{Sim.} \\
  & (\%) & (\%) & (\%) & (\%) & (\%) & (\%) & (\%) & (\%) & (\%) & (\%) & (\%) & (\%) & (\%) & (\%) & (\%) & (\%) & (\%) & (\%) & (\%) & (\%) & (\%) \\
  \hline 
  \textbf{Our Attack} & \textbf{1.2} & \textbf{4.8} & \textbf{97.6} & \textbf{4.7} & \textbf{11.9} & \textbf{93.5} & \textbf{4.7} & \textbf{14.2} & \textbf{87.4} & \textbf{4.8} & \textbf{13.8} & \textbf{88.1} & \textbf{5.0} & \textbf{15.6} & \textbf{83.6} & \textbf{7.1} & \textbf{18.2} & \textbf{81.9} & \textbf{7.3} & \textbf{18.8} & \textbf{84.2} \\
  Stats. Analysis & 6.7 & 22.0 & 82.0 & 20.0 & 51.0 & 25.1 & 18.3 & 51.8 & 6.2 & 22.4 & 52.5 & 7.4 & 19.3 & 51.4 & 19.1 & 17.9 & 46.3 & 6.9 & 13.2 & 34.2 & 56.8 \\
  Transferability & 77.8 & 99.6 & 0.0 & 57.8 & 100.0 & 0.0 & 53.4 & 100.0 & 0.0 & 47.4 & 86.2 & 0.0 & 61.8 & 96.1 & 0.0 & 31.9 & 87.2 & 0.0 & 49.9 & 92.8 & 0.0 \\
  Attack in~\cite{yangusenix2023} & 17.6 & 56.5 & 8.3 & 33.1 & 82.1 & 0.0 & 8.0 & 26.4 & 65.8 & 16.8 & 52.9 & 22.7 & 17.5 & 48.1 & 14.7 & 44.4 & 97.6 & 0.0 & 48.0 & 100.0 & 0.0 \\
  \hline
  \end{tabular}
  
  \centering
  \setlength{\tabcolsep}{.234em}
  \begin{tabular}{|ccc|ccc|ccc|ccc|ccc|ccc|ccc|ccc|}
  \hline
  \multicolumn{3}{|c|}{\textbf{P8}} & \multicolumn{3}{c|}{\textbf{P9}} & \multicolumn{3}{c|}{\textbf{P10}} & \multicolumn{3}{c|}{\textbf{P11}} & \multicolumn{3}{c|}{\textbf{P12}} & \multicolumn{3}{c|}{\textbf{P13}} & \multicolumn{3}{c|}{\textbf{P14}} & \multicolumn{3}{c|}{\textbf{P15}} \\
  \textbf{CER} & \textbf{WER} & \textbf{Sim.} & \textbf{CER} & \textbf{WER} & \textbf{Sim.} & \textbf{CER} & \textbf{WER} & \textbf{Sim.} & \textbf{CER} & \textbf{WER} & \textbf{Sim.} & \textbf{CER} & \textbf{WER} & \textbf{Sim.} & \textbf{CER} & \textbf{WER} & \textbf{Sim.} & \textbf{CER} & \textbf{WER} & \textbf{Sim.} & \textbf{CER} & \textbf{WER} & \textbf{Sim.} \\
  (\%) & (\%) & (\%) & (\%) & (\%) & (\%) & (\%) & (\%) & (\%) & (\%) & (\%) & (\%) & (\%) & (\%) & (\%) & (\%) & (\%) & (\%) & (\%) & (\%) & (\%) & (\%) & (\%) & (\%) \\
  \hline
  \textbf{8.4} & \textbf{21.7} & \textbf{80.5} & \textbf{8.8} & \textbf{25.8} & \textbf{77.7} & \textbf{9.5} & \textbf{25.8} & \textbf{77.7} & \textbf{12.5} & \textbf{34.5} & \textbf{63.9} & \textbf{13.3} & \textbf{33.9} & \textbf{50.7} & \textbf{13.3} & \textbf{34.1} & \textbf{51.1} & \textbf{14.2} & \textbf{44.6} & \textbf{34.4} & \textbf{15.9} & \textbf{44.4} & \textbf{19.7} \\
  28.1 & 62.8 & 0.0 & 27.6 & 57.7 & 0.0 & 24.0 & 61.4 & 4.1 & 29.1 & 70.1 & 0.0 & 31.7 & 70.3 & 0.0 & 45.0 & 91.6 & 0.0 & 38.6 & 83.4 & 0.0 & 54.5 & 96.8 & 0.0 \\
  44.3 & 100.0 & 0.0 & 47.7 & 90.6 & 0.0 & 47.7 & 85.7 & 0.0 & 54.3 & 94.3 & 0.0 & 20.1 & 51.0 & 25.3 & 91.0 & 100.0 & 0.0 & 14.3 & 33.9 & 59.3 & 47.9 & 94.0 & 0.0 \\
  32.7 & 78.9 & 0.0 & 30.5 & 64.3 & 0.0 & 14.9 & 37.2 & 50.6 & 16.2 & 41.9 & 36.7 & 26.6 & 67.5 & 0.0 & 20.0 & 53.6 & 5.5 & 31.6 & 91.0 & 0.0 & 62.4 & 99.2 & 0.0 \\
  \hline
  \end{tabular}
  \vspace{-0.05in}
  \caption{Attack performance for all 15 participants (P1-15).}
  \vspace{-0.2in}
  \label{tab:all_participants}
\end{table*}

\vspace{-0.1in}
\subsection{Performance across Users}
\label{subsec:evaluser}

To assess the effect of individual typing styles on the proposed attack, we recruited 15 participants to our user study (age: 21-53,  7 males and 8 females).  In Table \ref{tab:all_participants}, we report the performance of our \MakeLowercase{\typeI{}} ($\mathbf{H}_{org}$) attack against each participant. Overall, the attack is highly effective -- for 13 out of 15 participants, our attack accurately recognizes 86\%-98\% of typed keys, and the recovered text retains 50\%-98\% of the original content's meaning.

\para{The successful cases (P1-P13).} After closely observing  these users, we identified two distinct typing patterns that increase users' vulnerability to our attack. First, they type with prominent motion,  so the hand tracking sensors on the VR headset can capture sufficient amounts of finger movements. The hand tracking accuracy is hardware dependent and does vary across users. Second, their body posture and hand orientation minimize self-occlusion across their fingers, meaning the joints are not blocking each other from the view of the hand tracker (located on their VR headset). When a joint is obscured, the joint's coordinates are often inaccurately estimated, leading to random tracking noise.

\begin{figure}[h]
  \vspace{-0.1in}
  \centering
    \includegraphics[width=0.92\linewidth]{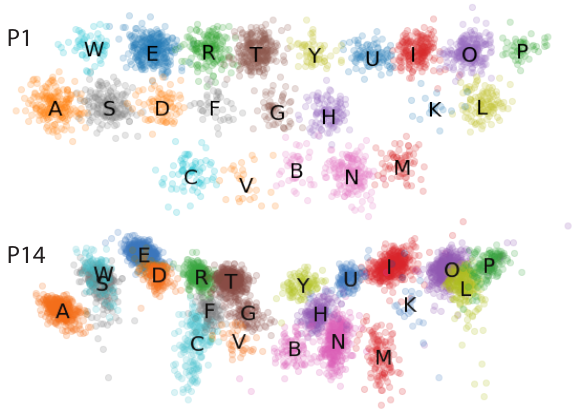}  
    \vspace{-0.05in}
    \caption{P1 and P14's keystroke pressing locations tracked by the VR headset. The locations are based on ground truth keystroke timestamps and pressing fingertips.}
    \label{fig:xz_error_per_letter}
    \vspace{-0.1in}
\end{figure}

\para{The unsuccessful case (P14 and P15).} Our attack is less effective on P14 and P15, due to high hand tracking noise.  For P14, the VR hand tracker is unable to track the pressing fingertip location (i.e., the $x$ and $z$ values) accurately. To illustrate this, Figure~\ref{fig:xz_error_per_letter} compares P14's keystroke pressing locations reported by the VR tracker to those of P1.  We see that P14 has many more overlapping keystrokes than P1, making it hard to create reliable labels. 

For P15, the major issue is the height data ($y$) collected by the VR hand tracker. This is because, rather than having the keystroking finger much closer to the keyboard than the other fingers, P15's fingers stay at roughly the same height level when pressing keys.  As a result, the tracking data is particularly noisy on the $y$ axis, making it hard to reliably detect keystroke events and the pressing finger.

\vspace{-0.1in}
\subsection{Comparison to Other Solutions}
\label{subsec:comparison}
\vspace{-0.05in}

We compare our attack design to baseline solutions discussed in \S\ref{subsec:explore} and related work~\cite{yangusenix2023} discussed in \S\ref{sec:designInsights}. Overall, our attack design significantly outperforms these solutions.  

\para{Our attack design vs. baselines.} We compare our approach with two baseline approaches discussed in \S\ref{subsec:explore}. The first baseline is {\em unsupervised inference via statistical analysis}, which takes three sequential steps of detecting keystrokes, identifying pressing fingertips, and recognizing keys. These are the same three steps used by our attack design to generate initial labels but this baseline does not involve any DNN models. The second baseline is {\em attack by transferability}, where for each target, we use ground truth labels on all 14, non-target participants to train a DNN keystroke classifier, and hope it can also recognize the target's keystrokes. Note that in this implementation of transferability-based attacks, we assume perfect keystroke detection on the target user. Thus, the result represents the best case of transferability-based attacks.

The results in Table \ref{tab:all_participants} show that our attack design significantly outperforms the two baselines. Compared to the {\em unsupervised inference via statistical analysis}, our approach curates labeled training data for training transformer-based DNN models, which can better capture the intrinsic features of the noisy training samples. The second baseline {\em attack by transferability} failed largely because human typing behaviors are highly user-specific and do not transfer to others~\cite{yangusenix2023}. Our approach overcomes this challenge by utilizing the target's own data to learn their specific typing behaviors. 

\para{Our attack design vs. \cite{yangusenix2023}.} 
We obtain the source code from the authors of \cite{yangusenix2023} and follow the training method described in \cite{yangusenix2023}. To train and test the CNN classifiers used in \cite{yangusenix2023}, we convert the telemetry frames to images. For a fair comparison, we apply the same conversion method used by our approach (see \S\ref{sec:hmm}).

The results in Table \ref{tab:all_participants} show that our attack design significantly outperforms that of~\cite{yangusenix2023}. As discussed in \S\ref{sec:designInsights}, the amount of noise in our attack data is significantly higher than that of~\cite{yangusenix2023}, rendering the method of~\cite{yangusenix2023} ineffective. Our design addresses this challenge by deploying multiple task-specific transformers where the self-attention mechanism allows these models to extract key patterns from noisy data. 

\vspace{-0.1in}
\subsection{Attack Complexity}
\label{subsec:evaltime}
\vspace{-0.05in}
We measure attack complexity by the amount of time required to produce the recovered text after the typing session stops. We run our end-to-end attack pipeline on a standard server with an AMD EPYC 7313P 16-Core processor, and train all the models on a single NVIDIA A40 GPU. The overall runtime of our attack is 423 minutes for a 500-word, 19-minute typing session. The majority of time was spent on the training of transformer-based keystroke detector (79 mins), the finger identifier (64 mins), and the CNN-based keystroke classifier (259 mins). The other components only take 21 minutes in total. Specifically, clustering \& HMM take 3 minutes and spell checking takes 11 minutes. 
\vspace{-0.1in}
\section{Defenses}
\label{sec:defense}
\vspace{-0.05in}

Our study shows that by either intercepting the telemetry data used to render a target's avatar or simply observing rendered hand movements of their avatar, an attacker in the same virtual environment can successfully recover the text physically typed by the VR user. Untreated, these attacks can cause significant damage to users. Given this threat, it is important for VR platform and app developers to exercise caution and implement adequate safeguards during the development process. We discuss several defense options below and their impact on the attacks and the VR experience. 

\para{Defense 1: limiting access to telemetry (hand tracking) data.} The first type of defense seeks to minimize the chance of leaking sensitive handpose data ($\mathbf{H}_{org}$ and $\mathbf{T}_{3d}$) to attackers.  After detecting the user is typing, the VR system can either ban access to its hand tracking API by any application, put a significant limit on the query frequency, or largely reduce the sampling rate of the hand tracking module. For example, by reducing the sampling rate from the default 60fps to 15fps, the defense can increase attack WER from 14.2\% to 38\%. When further reduced to 6fps, the attack becomes ineffective.

\para{Defense 2: adding noise to telemetry data.} Another type of defense is to add noise to the 3D hand tracking data captured by the wearer's headset (i.e.,  $\mathbf{H}_{org}$). In particular, by perturbing the depth ($y$) value of $\mathbf{H}_{org}$, one can effectively confuse the attacker at the stages of keystroke detection and finger identification, preventing them from identifying the keystroke events and/or the pressing finger. In our design, we choose to add zero-mean Gaussian noise to the $y$ value. This is because our empirical measurements show that the noise naturally present in the hand tracking data follows a Gaussian distribution.  Adding zero-mean Gaussian noise on top makes it hard for the attacker to denoise the data. 

Table~\ref{tab:add_noise} summarizes the attack performance on the original telemetry data ($\mathbf{H}_{org}$) before and after applying this defense. We consider two noise levels: {\em moderate}, where we add zero-mean Gaussian noise with an STD of  0.3 $\times$ key width, and {\em high}, where the STD rises to 0.5 $\times$ key width. We see that adding a moderate level of noise can already disrupt the attack, raising CER from 4.7\% to 20.1\%, WER from 14.2\% to 55.4\%, and dropping semantic similarity to 14\%. Under a high level of noise, WER further rises to 71.4\% and semantic similarity drops to 0\%.  We also visually inspect the target's avatar under both noise levels. The avatar still displays normal typing behavior under moderate noise, while the typing speed appears to be considerably accelerated under high noise. 

\begin{table}[h]
\renewcommand{\arraystretch}{1.005}
\centering
\setlength{\tabcolsep}{0.9em}
\begin{tabular}{|l|rrr|}
\hline
\textbf{Noise Level} & \textbf{CER} & \textbf{WER} & \textbf{Similarity} \\
 (STD) & (\%) & (\%) & (\%)\\
\hline
No noise & 4.7 & 14.2 & 87.4 \\ 
Moderate noise: 0.3 $\times$ Key Width & 20.1 & 55.4 & 14.1 \\
High noise: 0.5 $\times$ Key Width & 26.6 & 71.4 &  0.0 \\
\hline
\end{tabular}
\vspace{-0.05in}
\caption{Attack performance after perturbing the depth axis of $\mathbf{H}_{org}$ using zero-mean Gaussian noise.}
\vspace{-0.1in}
\label{tab:add_noise}
\end{table}

\para{Impact on VR immersiveness.}  As expected, both defenses could impact the VR experience, because the user's physical hand motion is no longer synchronized with their avatar. Here the user can choose to run the defense directly on telemetry data collected by the VR headset, or only when the collected telemetry data leaves the VR headset.   The first option provides a more thorough protection but affects how the target interacts with the VR environment. In particular, the target user may not see their correct typing movements inside the VR. The second option only affects the avatar observed by other VR users, but cannot resist attacks that can access the headset's telemetry data. Clearly, this represents an inherent tradeoff between usability and security.
\vspace{-0.1in}
\section{Conclusion}
\vspace{-0.1in}
Our study is the first to explore the possibility of keystroke inference attacks in a shared virtual reality environment. In this scenario, an adversary VR user can reconstruct text typed by another user by observing their avatar. Unlike prior work, our attack does not require physical observation of the target, only a noisy VR representation of their avatar's hand movements. This avatar-based attack highlights the tension between immersive experiences and personal/information security in VR. While adding noise or reducing tracking frequency can reduce the attack's effectiveness, it also impacts the immersive experience to varying degrees. This raises the important question of how to design VR systems that balance the user's desire for immersive and engaging experiences with the need to protect their personal and information security. 

\para{Limitations and Future Work.} The major limitation of our attack pipeline is that we ultimately choose a 3D-CNN model for our keystroke recognizer because the self-labeled training data for this component is very limited and imbalanced. The 3D-CNN works adequately with limited training data as it has already been pretrained on a large gesture recognition dataset~\cite{materzynska2019jester}. We were not able to find pretrained transformer models or large datasets with telemetry data for both hands. Nevertheless, as transformers are being successfully applied to non-NLP domains, it is reasonable to expect that pretrained models in the future would make our attack more effective.

\vspace{-0.1in}
\section*{Acknowledgements}
\vspace{-0.1in}
We thank our anonymous reviewers and shepherd for their insightful
feedback. This work is supported in part by NSF grants CNS-1949650,
CNS-1923778, CNS-2241303, and the DARPA GARD program. Opinions, findings, and conclusions or recommendations expressed in this material are those of the authors and do not necessarily reflect the views of any funding agencies.

\bibliographystyle{plain}
\bibliography{ref}

\begin{thebibliography}{10}

\bibitem{keyboardvr}
The {Logitech} {K830} keyboard and typing in {VR}.
\newblock
  \url{https://medium.com/xrlo-extended-reality-lowdown/the-logitech-k830-keyboard-and-typing-in-vr-556e2740c48d}.

\bibitem{keyboardvrhtc}
Type in {VR} with {Logitech}'s keyboard for {HTC} {Vive}.
\newblock
  \url{https://vrscout.com/news/type-in-vr-logitech-keyboard-kit-for-vive/}.

\bibitem{amini2022self}
Massih-Reza Amini, Vasilii Feofanov, Loic Pauletto, Emilie Devijver, and Yury
  Maximov.
\newblock Self-training: A survey.
\newblock {\em arXiv}, 2022.

\bibitem{visionpro}
Apple.
\newblock Vision {Pro}.
\newblock \url{https://www.apple.com/apple-vision-pro/}.

\bibitem{Arafat2021}
Abdullah~Al Arafat, Zhishan Guo, and Amro Awad.
\newblock {VR-Spy}: A side-channel attack on virtual key-logging in vr
  headsets.
\newblock In {\em Proc. of IEEE VR}, 2021.

\bibitem{arazo2019unsupervised}
Eric Arazo, Diego Ortego, Paul Albert, Noel O'Connor, and Kevin McGuinness.
\newblock Unsupervised label noise modeling and loss correction.
\newblock In {\em Proc. of ICML}, 2019.

\bibitem{bahdanau2014neural}
Dzmitry Bahdanau, Kyunghyun Cho, and Yoshua Bengio.
\newblock Neural machine translation by jointly learning to align and
  translate.
\newblock {\em arXiv}, 2014.

\bibitem{clearshot}
Davide Balzarotti, Marco Cova, and Giovanni Vigna.
\newblock Clearshot: Eavesdropping on keyboard input from video.
\newblock In {\em Proc. of IEEE S\&P}, 2008.

\bibitem{banerjee2012biometric}
Salil~P. Banerjee and Damon~L. Woodard.
\newblock Biometric authentication and identification using keystroke dynamics:
  A survey.
\newblock {\em Journal of Pattern Recognition Research}, 7(1), 2012.

\bibitem{baum1972inequality}
Leonard~E Baum.
\newblock An inequality and associated maximization technique in statistical
  estimation for probabilistic functions of markov processes.
\newblock {\em Inequalities}, 3(1):1--8, 1972.

\bibitem{mediapipe}
Valentin Bazarevsky and Fan Zhang.
\newblock On-device, real-time hand tracking with {MediaPipe}.
\newblock
  \url{https://ai.googleblog.com/2019/08/on-device-real-time-hand-tracking-with.html},
  2021.

\bibitem{vrphykeyboard}
Doug~A. Bowman.
\newblock Embracing physical keyboards for virtual reality.
\newblock {\em Computer}, 53(09):9--10, 2020.

\bibitem{perspectiveprojection}
Wayne Brown.
\newblock Perspective projections.
\newblock
  \url{http://learnwebgl.brown37.net/08_projections/projections_perspective.html}.

\bibitem{keystrokchi15}
Daniel Buschek, Alexander~De Luca, and Florian Alt.
\newblock Improving accuracy, applicability and usability of keystroke
  biometrics on mobile touchscreen devices.
\newblock In {\em Proc. of CHI}, 2015.

\bibitem{wifi2018}
Yunfang Chen, Yihong Zhu, Hao Zhou, Wei Chen, and Wei Zhang.
\newblock Enhanced keystroke recognition based on moving distance of keystrokes
  through {WiFi}.
\newblock In {\em Proc. of NSS}, 2018.

\bibitem{chu2020expressive}
Hang Chu, Shugao Ma, Fernando~De la~Torre, Sanja Fidler, and Yaser Sheikh.
\newblock Expressive telepresence via modular codec avatars.
\newblock In {\em Proc. of ECCV}, 2020.

\bibitem{Enron}
William~W. Cohen.
\newblock Enron email dataset.
\newblock \url{https://www.cs.cmu.edu/~enron/}, 2015.

\bibitem{copyleaks}
CopyLeaks.
\newblock Plagiarism checker api - integrate ai powered api, copyleaks.
\newblock \url{https://api.copyleaks.com/}.

\bibitem{hermann2015teaching}
Karl Moritz~Hermann et~al.
\newblock Teaching machines to read and comprehend.
\newblock {\em Advances in neural information processing systems}, 2015.

\bibitem{han2020megatrack}
Shangchen~Han et~al.
\newblock {MEgATrack}: Monochrome egocentric articulated hand-tracking for
  virtual reality.
\newblock {\em ACM Transactions on Graphics}, 39, 2020.

\bibitem{zhu2021converting}
Yitan~Zhu et~al.
\newblock Converting tabular data into images for deep learning with
  convolutional neural networks.
\newblock {\em Scientific Reports}, 2021.

\bibitem{wercompute}
Hugging Face.
\newblock {WER} - a hugging face space by evaluate-metric.
\newblock \url{https://huggingface.co/spaces/evaluate-metric/wer}.

\bibitem{Feit2016_how_we_type}
Anna~Maria Feit, Daryl Weir, and Antti Oulasvirta.
\newblock {How We Type}: Movement strategies and performance in everyday
  typing.
\newblock In {\em Proc. of CHI}, 2016.

\bibitem{stereoDepthEstimation}
Sanja Fidler.
\newblock Stereo epipolar geometry for general cameras.
\newblock
  \url{http://www.cs.toronto.edu/~fidler/slides/2021Winter/CSC420/lecture13.pdf},
  2021.

\bibitem{garrido2023sok}
Gonzalo~Munilla Garrido, Vivek Nair, and Dawn Song.
\newblock {SoK}: Data privacy in virtual reality.
\newblock {\em arXiv}, 2023.

\bibitem{giaretta2022security}
Alberto Giaretta.
\newblock Security and privacy in virtual reality -- a literature survey.
\newblock {\em arXiv}, 2022.

\bibitem{googledocs}
Jayakumar Hoskere.
\newblock Everyday {AI}: Beyond spell check, how google docs is smart enough to
  correct grammar | google cloud blog.
\newblock
  \url{https://cloud.google.com/blog/products/g-suite/everyday-ai-beyond-spell-check-how-google-docs-is-smart-enough-to-correct-grammar}.

\bibitem{howard2019avatars}
Mike Howard.
\newblock Avatars: The art and science of social presence.
\newblock
  \url{https://www.meta.com/blog/quest/avatars-the-art-and-science-of-social-presence/}.

\bibitem{immersed}
immersed.
\newblock Empower work with spatial computing.
\newblock \url{https://immersed.com}.

\bibitem{jank2006algorithm}
Wolfgang Jank.
\newblock The {EM} algorithm, its randomized implementation and global
  optimization: Some challenges and opportunities for operations research.
\newblock {\em Perspectives in operations research}, 2006.

\bibitem{emccs21}
Wenqiang Jin, Srinivasan Murali, Huadi Zhu, and Ming Li.
\newblock Periscope: A keystroke inference attack using human coupled
  electromagnetic emanations.
\newblock In {\em Proc. of ACM CCS}, 2021.

\bibitem{rotationmatrix}
Coding Labs.
\newblock World, view and projection transformation matrices.
\newblock
  \url{http://www.codinglabs.net/article_world_view_projection_matrix.aspx}.

\bibitem{windtalker}
Mengyuan Li, Yan Meng, Junyi Liu, Haojin Zhu, Xiaohui Liang, Yao Liu, and
  Na~Ruan.
\newblock When {CSI} meets public {WiFi}: Inferring your mobile phone password
  via {WiFi} signals.
\newblock In {\em Proc. of ACM CCS}, 2016.

\bibitem{lim2020revisiting}
John Lim, True Price, Fabian Monrose, and Jan-Michael Frahm.
\newblock Revisiting the threat space for vision-based keystroke inference
  attacks.
\newblock In {\em Proc. of ECCV}, 2020.

\bibitem{spidermon}
Kang Ling, Yuntang Liu, Ke~Sun, Wei Wang, Lei Xie, and Qing Gu.
\newblock {SpiderMon}: Towards using cell towers as illuminating sources for
  keystroke monitoring.
\newblock In {\em Proc. of IEEE INFOCOM}, 2020.

\bibitem{lingcns2019}
Zhen Ling, Zupei Li, Chen Chen, Junzhou Luo, Wei Yu, and Xinwen Fu.
\newblock I know what you enter on {Gear} {VR}.
\newblock In {\em Proc. of {CNS}}, 2019.

\bibitem{Luo2022}
Shiqing Luo, Xinyu Hu, and Zhisheng Yan.
\newblock {HoloLogger}: Keystroke inference on mixed reality head mounted
  displays.
\newblock In {\em Proc. of IEEE VR}, 2022.

\bibitem{materzynska2019jester}
Joanna Materzynska, Guillaume Berger, Ingo Bax, and Roland Memisevic.
\newblock The jester dataset: A large-scale video dataset of human gestures.
\newblock In {\em Proc. of ICCV Workshops}, 2019.

\bibitem{tapid2021}
Manuel Meier, Paul Streli, Andreas Fender, and Christian Holz.
\newblock {TapID}: Rapid touch interaction in virtual reality using wearable
  sensing.
\newblock In {\em Proc. of IEEE VR}, 2021.

\bibitem{meister2021companies}
Jeanne~C. Meister.
\newblock How companies are using {VR} to develop employees' soft skills.
\newblock {\em Harvard Business Review}, 2021.

\bibitem{horizon-workrooms}
Meta.
\newblock Meta {Horizon} {Workrooms}.
\newblock \url{https://forwork.meta.com/horizon-workrooms}.

\bibitem{pwc2022survey}
PwC.
\newblock {PwC} 2022 us metaverse survey.
\newblock
  \url{https://www.pwc.com/us/en/tech-effect/emerging-tech/metaverse-survey.html},
  2022.

\bibitem{ispy}
Rahul Raguram, Andrew~M. White, Dibyendusekhar Goswami, Fabian Monrose, and
  Jan-Michael Frahm.
\newblock {IS}py: Automatic reconstruction of typed input from compromising
  reflections.
\newblock In {\em Proc. of ACM CCS}, 2011.

\bibitem{SeleniumHQ}
SeleniumHQ.
\newblock {SeleniumHQ}/selenium: A browser automation framework and ecosystem.
\newblock \url{https://github.com/SeleniumHQ/selenium}.

\bibitem{sharma2019deepinsight}
Alok Sharma, Edwin Vans, Daichi Shigemizu, Keith~A. Boroevich, and Tatsuhiko
  Tsunoda.
\newblock {DeepInsight}: A methodology to transform a non-image data to an
  image for convolution neural network architecture.
\newblock {\em Scientific Reports}, 2019.

\bibitem{dstanet_accv2020}
Lei Shi, Yifan Zhang, Jian Cheng, and Hanqing Lu.
\newblock Decoupled spatial-temporal attention network for skeleton-based
  action-gesture recognition.
\newblock In {\em Proc. of ACCV}, 2020.

\bibitem{beware}
Diksha Shukla, Rajesh Kumar, Abdul Serwadda, and Vir~V. Phoha.
\newblock Beware, your hands reveal your secrets!
\newblock In {\em Proc. of ACM CCS}, 2014.

\bibitem{openpose}
Tomas Simon, Hanbyul Joo, Iain Matthews, and Yaser Sheikh.
\newblock Hand keypoint detection in single images using multiview
  bootstrapping.
\newblock In {\em Proc. of CVPR}, 2017.

\bibitem{song2001timing}
Dawn~Xiaodong Song, David Wagner, and Xuqing Tian.
\newblock Timing analysis of keystrokes and timing attacks on {SSH}.
\newblock In {\em Proc. of USENIX Security Symposium}, 2001.

\bibitem{Meteriz2022}
\"{U}lk\"{u} Meteriz-Yildiran, Necip~Fazil Yildiran, Amro Awad, and David
  Mohaisen.
\newblock A keylogging inference attack on air-tapping keyboards in virtual
  environments.
\newblock In {\em Proc. of IEEE VR}, 2022.

\bibitem{vaswani2017attention}
Ashish Vaswani, Noam Shazeer, Niki Parmar, Jakob Uszkoreit, Llion Jones,
  Aidan~N. Gomez, {\L}ukasz Kaiser, and Illia Polosukhin.
\newblock Attention is all you need.
\newblock {\em Proc. of NeurIPS}, 2017.

\bibitem{QoMEX2020}
Jan-Niklas Voigt-Antons, Tanja Kojic, Danish Ali, and Sebastian Möller.
\newblock Influence of hand tracking as a way of interaction in virtual reality
  on user experience.
\newblock In {\em Proc. of QoMEX}, 2020.

\bibitem{occulusimu}
Oculus VR.
\newblock Stereo-based calibration apparatus.
\newblock \url{https://patents.justia.com/patent/9805512}.

\bibitem{vspatial}
vSpatial.
\newblock The workspace of the future is here.
\newblock \url{https://www.vspatial.com/xr}.

\bibitem{xie2017aggregated}
Saining Xie, Ross Girshick, Piotr Doll{\'a}r, Zhuowen Tu, and Kaiming He.
\newblock Aggregated residual transformations for deep neural networks.
\newblock In {\em Proc. of CVPR}, 2017.

\bibitem{seeingdouble}
Yi~Xu, Jared Heinly, Andrew~M. White, Fabian Monrose, and Jan-Michael Frahm.
\newblock Seeing double: Reconstructing obscured typed input from repeated
  compromising reflections.
\newblock In {\em Proc. of ACM CCS}, 2013.

\bibitem{yangton2022}
Edwin Yang, Song Fang, IanMarkwood, Yao Liu, Shangqing Zhao, Zhuo Lu, and
  Haojin Zhu.
\newblock Wireless training-free keystroke inference attack and defense.
\newblock {\em IEEE/ACM Transactions on Networking}, 30(4):1733--1748, 2022.

\bibitem{yangusenix2023}
Zhuolin Yang, Yuxin Chen, Zain Sarwar, Hadleigh Schwartz, Ben~Y. Zhao, and
  Haitao Zheng.
\newblock Towards a general video-based keystroke inference attack.
\newblock In {\em Proc. of USENIX Security Symposium}, 2023.

\bibitem{blind1}
Qinggang Yue, Zhen Ling, Xinwen Fu, Benyuan Liu, Kui Ren, and Wei Zhao.
\newblock Blind recognition of touched keys on mobile devices.
\newblock In {\em Proc. of ACM CCS}, 2014.

\bibitem{zhang2017mixup}
Hongyi Zhang, Moustapha Cisse, Yann~N. Dauphin, and David Lopez-Paz.
\newblock mixup: Beyond empirical risk minimization.
\newblock {\em arXiv}, 2017.

\bibitem{zhu2005semi}
Xiaojin Zhu.
\newblock Semi-supervised learning literature survey.
\newblock {\em University of Wisconsin -- Madison}, 2008.

\bibitem{zhuang2005}
Li~Zhuang, Feng Zhou, and J.~D. Tygar.
\newblock Keyboard acoustic emanations revisited.
\newblock In {\em Proc. of ACM CCS}, 2005.

\end{thebibliography}
\section*{Appendix}

\subsection{Clustering}
\label{appendix:clusteringdetails}

\para{Detecting and clustering thumb generated keystrokes.} We observe that most users have very subtle motions with regards to typing keys with their thumbs (usually the space key). Therefore, our motion and displacement based techniques do not work well for thumb tip data. However, we are still able to detect thumb-based keystrokes via the movement of the other four fingers, because they have a distinct movement pattern when the thumb is used to press a key. Specifically, the four fingers move in tandem much more as compared to when a non-thumb finger is used for a keystroke. So if the fingers move together, the variance in their relative displacement is low and we use this signal to mark detected keystrokes as being thumb-generated.

However, this process is not very accurate and we follow the approach laid out by~\cite{yangusenix2023} to deal with thumb generated keystrokes. Despite having some confidence that a keystroke was thumb-generated, we treat it as if it was a non-thumb keystroke i.e., we proceed by including it in the clustering process by using the coordinates of the non-thumb key identified as having generated that keystroke. The only difference is that we do not let these keystrokes be clustered in the regular non-thumb key clusters. Rather, we let these keys form their own clusters and let the HMM decide which of these clusters correspond to thumb keystrokes, i.e.,  space keys and which are non-thumb keystrokes.

\para{Labeling consistency within each cluster.} If the clustering was perfectly clean, we would expect each cluster to contain touchpoints of the same key and the HMM would most often label the clusters as such. However, given the errors in our clustering, the HMM often assigns different labels to keystrokes within the same cluster as it is not bound to assign a one-to-one mapping between a cluster's ID and its associated label. Rather, it relies on its learned parameters to decide the most likely labels at each time step. While the HMM can often assign the correct label to an incorrectly clustered keystroke, we find that minority labels within a cluster are unreliable. Therefore, we only select labels from the most dominant class in each cluster.

\para{Cross-cluster label compatibility.} By inspecting the majority label assigned by the HMM to the keystrokes in each cluster, we are further able to identify errors by previous modules in our pipeline. Specifically, we can identify errors in finger identification and the presence of non-keystroke events in our pipeline by doing a cross-cluster label consistency check. Intuitively, if two clusters have the same majority label, they should be close together in the clustering space since the clustering is spatial and each key has a fixed location on the keyboard. Therefore, if two clusters are far apart and have been assigned the same majority label, one of them cannot be legitimate.  Following this observation, we sort such clusters by size and remove all but the largest one from our training set for the keystroke classifier.

\para{Detecting backspaces} For accurate keystroke inference, it is imperative that we detect backspaces with high precision as failure to do so leads to high rates of error accumulation. We use two heuristics to identify backspaces among the set of detected keystrokes. We assume that the backspace is one of the edge keys on the keyboard and secondly, it is often pressed multiple times together to fix typos. As a result, we label the corner cluster on our touchpoint map with the highest ratio of consecutive keystrokes associated with it. For every keystroke in the backspace cluster, we remove the previous non-backspace detected keystroke for it. After this step, the backspace cluster is not used anymore in the attack pipeline.

\vspace{-0.075in}
\subsection{Hidden Markov Models (HMM)}
\vspace{-0.075in}
\label{appendix:hmm}

The output of the clustering module gives us a sequence of cluster IDs where each element in the sequence corresponds to a keystroke in the order that it occurred. We want to map each element in the sequence to 29 keys (period, comma, space key, and the 26 letters of the English alphabet) and a sequence tagging algorithm such as the HMM can perform this mapping. We are interested in obtaining the actual keys being typed by the victim but are unable to directly observe them. However, we have obtained a sequence of cluster IDs that are causally related to the actual keys being typed. HMMs can model this causal relationship between the observed and unobserved events and then infer the unobserved events. 

Formally, HMMs are characterized by two sets of parameters \textbf{A} and  called the Transition and Emission matrices respectively. Matrix \textbf{A} models the probability of moving from one hidden state to another whereas the matrix \textbf{B} models the probabilities of hidden states producing the observed states. In our case, \textbf{A} models the actual \textit{N=29} keys and is a \textit{N X N} matrix whereas \textbf{B} models the \textit{M < 50} cluster IDs being produced by the alphabets and is therefore a \textit{N X M} matrix. Probabilities in both matrices satisfy the Markov assumption. Inspired by~\cite{yangusenix2023}, we compute \textbf{A} by empirically estimating the transition probabilities in the English language by using 40,000 sentences from the CNN/DailyMail dataset~\cite{hermann2015teaching}. Matrix \textbf{B} is learned using a variant of the Expectation-Maximization (E-M)~\cite{baum1972inequality} algorithm given \textbf{A} and the cluster sequence. Given \textbf{A, B}, and the cluster sequence, the HMM infers the most probable hidden state sequence, i.e., the typed keys. 

However, since the E-M algorithm performs local optimization~\cite{jank2006algorithm} it often converges to local minimas. To deal with this, we run multiple iterations of the HMM and select the model that produces the most number of high-confidence predicted labels using the consistency checks explained in Appendix~\ref{appendix:clusteringdetails}.

\begin{table}[h]
\vspace{-0.05in}
\renewcommand{\arraystretch}{1.005}
\footnotesize
\centering
\setlength{\tabcolsep}{0.444em}
\begin{tabular}{|l|l|rrr|rrr|}

\hline
\multicolumn{1}{|l|}{\multirow[c]{3}{*}{\parbox{0.8cm}{\textbf{Attack Data}}}} & \multicolumn{1}{l|}{\multirow[c]{3}{*}{\parbox{1.2cm}{\textbf{Attacker's PoV}}}} & \multicolumn{3}{c|}{\textbf{P1}} & \multicolumn{3}{c|}{\textbf{P2}} \\ 
 & & \textbf{CER} & \textbf{WER} & \textbf{Sim.} &\textbf{CER} & \textbf{WER} & \textbf{Sim.} \\
 & & (\%) & (\%) & (\%) & (\%) & (\%) & (\%) \\
\hline

$\mathbf{H}_{org}$ & -- & 1.2 & 4.8 & 97.6 & 9.5 & 25.8 & 77.7 \\
\hline 

\multirow[c]{3}{*}{$\mathbf{T}_{3d}(PoV)$} 
& $(0, 0)$ & 2.1 & 7.2 & 95.6 & 10.3 & 29.1 & 75.5 \\
& $(0, 45)$ & 1.7 & 5.0 & 97.8 & 9.9 & 27.9 & 67.3 \\
& $(45, 0)$ & 1.6 & 6.6 & 95.8 & 9.4 & 25.0 & 70.3 \\

\hline 

\multirow[c]{2}{*}{$\mathbf{T}_{2d}(PoV)$} 
& $(0, 0)$ & 56.9 & 96.8 & 0.0 & 32.5 & 77.9 & 0.0 \\
& $(0, 45)$ & 10.2 & 26.1 & 74.4 & 41.7 & 93.8 & 0.0 \\

\hline 

\multirow[c]{2}{*}{\parbox{1.3cm}{$\mathbf{T}_{2d}(0, 0)$ \& $\mathbf{T}_{2d}(PoV)$}}
& $(0, 45)$ & 1.9 &  7.4 &  96.4 & 10.0 & 26.9 & 69.9 \\
& $(15, 0)$ & 1.6 &  5.2 &  96.8 & 9.3 & 25.4 & 72.7 \\

\hline

\end{tabular}
\vspace{-0.05in}
\caption{Attack performance on two other participants' telemetry data. We include $\mathbf{H}_{org}$, $\mathbf{T}_{3d}$, $\mathbf{T}_{2d}$, and an extended \MakeLowercase{\typeIII{}}
attack using two $\mathbf{T}_{2d}$ at different camera $\pov$s.}
\label{tab:transformed_others}
\vspace{-0.05in}
\end{table}
\begin{table}[h]
    \vspace{-0.15in}
    \renewcommand{\arraystretch}{1.005}
    \centering
    \setlength{\tabcolsep}{0.8em}
    \begin{tabular}{|l|cc|cc|}
    \hline
     & \multicolumn{2}{c|}{\textbf{Keystroke Detector $\mathcal{D}$}} & \multicolumn{2}{c|}{\textbf{Keystroke Classifier $\mathcal{C}$}} \\
     & \textbf{Training} & \textbf{Inference} & \textbf{Training} & \textbf{Inference} \\
     & \textit{keystroking frames} & \textit{all frames} & \multicolumn{2}{c|}{\textit{keystroking frames}} \\
    \hline
    P1 & 2925 & 181394 & 1647 & 2868 \\
    P2 & 2677 & 184139 & 1540 & 2849 \\
    P3 & 3070 &  50716 & 1475 & 2903 \\
    P4 & 3009 &  77494 & 1576 & 2950 \\
    P5 & 3096 &  70017 & 1351 & 3027 \\
    P6 & 3034 & 215462 & 1248 & 2916 \\
    P7 & 3235 & 184145 & 1676 & 2894 \\
    P8 & 3381 & 126165 & 1405 & 2887 \\
    P9 & 3023 & 121333 & 1257 & 2839 \\
    P10 & 3108 &  76549 & 1338 & 3026 \\
    P11 & 3078 &  79432 & 1341 & 2866 \\
    P12 & 3200 &  65410 & 1412 & 2950 \\
    P13 & 3194 &  37902 & 1233 & 2858 \\
    P14 & 3078 &  47422 & 1369 & 3042 \\
    P15 & 3274 &  36733 & 1333 & 2896 \\
    \hline
    \end{tabular}\vspace{-0.05in}
    \caption{For each of the 15 participants, the amount of training data, in terms of
        number of telemetry frames, used to train
        the DNNs (i.e., the keystroke detector $\mathcal{D}$ and the
        keystroke classifier $\mathcal{C}$), and the amount of
        inference data being fed into the trained models.  When
        training $\mathcal{D}$, we set the
        self-labeled 
        keystroking frames as positive samples in the training
        dataset, and extract the same amount of non-keystroking
        frames as the negative samples in the training dataset.}
\vspace{-0.15in}
\label{tab:dataset}
\end{table}

\end{document}